\documentclass[12pt]{article}

\usepackage{amsmath,amsfonts,epsfig,color,latexsym}

\newcommand{\be}{\begin{equation}}
\newcommand{\ee}{\end{equation}}
\newcommand{\bea}{\begin{eqnarray}}
\newcommand{\eea}{\end{eqnarray}}

\renewcommand{\theequation}{\thesection.\arabic{equation}}

\let\newsection=\section
\renewcommand{\section}{\setcounter{equation}{0}\newsection}

\begin{document}

\begin{flushright}
hep-th/0410173\\
BROWN-HET-1423
\end{flushright}
\vskip.5in

\begin{center}

{\LARGE\bf High energy QCD from Planckian 
scattering in AdS and the Froissart bound}
\vskip 1in
\centerline{\Large Kyungsik Kang and Horatiu Nastase}
\vskip .5in

\end{center}
\centerline{\large Brown University}
\centerline{\large Providence, RI, 02912, USA}

\vskip 1in

\begin{abstract}
{\large
We reanalyze high energy QCD scattering regimes from scattering in cut-off 
AdS via gravity-gauge dualities (a la Polchinski-Strassler). We look at 
't Hooft scattering, Regge behaviour and black hole creation in AdS. 
Black hole creation in the gravity dual is analyzed via gravitational 
shockwave collisions. We prove the saturation of the QCD Froissart 
unitarity bound,
corresponding to the creation of black holes of AdS size, as suggested 
by Giddings.

}

\end{abstract}

\newpage

\section{Introduction}

Gravity-gauge dualities have been an important tool in getting information 
about nonperturbative Yang-Mills theories, since the original work of 
Maldacena \cite{malda} (see also \cite{bmn}). But the contact with real 
high energy experiments has been lacking, partly because of the absence of a 
gravity dual of QCD. However, in the work of Polchinski and Strassler 
\cite{ps} (and the later \cite{pstwo} dealing with deep inelastic scattering)
it was shown that one can derive a lot of information from just a simple 
gravity dual model of AdS cut-off in the IR (where the mass gap is related 
to a modification of the geometry) and maybe in the UV, in other words 
the 2-brane Randall-Sundrum model \cite{rs}. 

High energy QCD scattering of colourless objects (e.g. glueballs)
is thus related to scattering in the cut-off 
AdS via convoluting the scattering amplitude with AdS wavefunctions. This 
simple model still allows one to get a lot of information about QCD at high 
energy, when the IR modifications due to the mass gap are not so important. 

Giddings \cite{gid} took this proposal further and analyzed the scattering 
inside AdS in the extreme inelastic case, when black holes are being formed. 
He analyzed the black hole formation using a simple model, devised in 
\cite{gt,dl} (see also earlier work in \cite{bf}), 
the cross section for black hole formation being equated with the 
geometric area of the black hole horizon of mass equal to the center of 
mass energy ($\sigma= \pi r_H^2$, $r_H=r_H(\sqrt{s})$). At moderate energies,
the black holes being formed can be taken to be in flat space, but at 
higher energies, the size of AdS becomes important. Giddings argued, by 
calculating the Newton-like potential (linearized Einstein gravity), that the 
horizon of a black hole formed on the RS IR brane will be such as to give 
a saturation of the Froissart unitarity bound for the cross section:
\be
h_{00; lin}\sim G_4 \sqrt{s} \frac{e^{-M_1 r}}{r}\Rightarrow r_H\sim 
\frac{1}{M_1}ln (G_4 \sqrt{s} M_1)\Rightarrow \sigma \simeq \pi r_H^2 
\sim (\frac{1}{M_1}ln (\sqrt{s}))^2
\ee

But there are a lot of uncertainties about this calculation. In a previous 
paper \cite{kn}, we addressed some of these uncertainties, by looking at 
black hole formation via shockwave collisions (a technique first proposed in 
\cite{eg}), and estimated the maximum impact parameter $b_{max}$
for which a black hole is being formed, and calculated the minimum mass of 
the formed black hole. The calculation was done in flat $D>4$, as well as 
in the background of the one brane RS model (``alternative to 
compactification'' \cite{rstwo}), as was apropriate for the various
low $M_{Pl}$ scenarios \cite{add,aadd,rs}
in which one could detect black holes at accelerators.
We have set up the general formalism in a way to be used for the AdS and 
2-brane RS calculations needed for the QCD dual scattering. We also analyzed 
the effect of string ($\alpha '$) corrections to the black hole 
creation. Quantum corrections were also analyzed in a different way in 
\cite{ry,gr} and 't Hooft scattering inside AdS was analyzed using different
methods in \cite{aho}.
The advantage in this semiclassical formalism is that 
now $\sigma = \pi b_{max}^2$ is rigorous.

In this paper we will explore the consequences of the calculations in 
\cite{kn} for the high energy QCD scattering, and we will revisit some of the 
previous results to see if we can gain more insight. We will try to use 
consistently the Polchinski-Strassler set-up, in particular finding 
how to turn 
the classical scattering with black hole formation, happening at a certain 
point in the gravity dual, into an integration over the gravity dual. We will
use a simple black disk model to turn the classical scattering into an 
imaginary elastic amplitude for which we can use the P-S formalism.  
The most important 
piece of information learned in this paper will be that we are able to justify 
in a more rigorous way the appearence of the Froissart bound. Given the 
Polchinski- Strassler set-up, we will calculate the maximum impact parameter 
being formed in the scattering of two shockwaves, and thus get $\sigma = 
\pi b_{max}^2$. The importance of this formalism is that the shockwaves 
are exact solutions (not linearized ones), giving an advantage over the 
horizon calculation in \cite{gid}. 

We should note here that we are not attempting to define a general amplitude
(nor a S matrix) for a general scattering inside AdS. We will see in the 
next two subsections that for our purposes it is enough to assume the 
existence of a nonperturbative amplitude (and corresponding cross section) 
when the external legs are situated in a flat 4d slice of AdS space. We will 
define a simple black disk model for this amplitude that mimics the 
(as yet unknown and undefined) nonperturbative amplitude by giving the same 
cross section. For external legs living in a flat 4d slice, one can certainly 
define amplitudes and cross sections in the usual manner. 

The paper is organized as follows. In section 2 we review the formalism 
of high energy scattering in QCD from scattering in cut-off AdS. In 
section 2.1 we show how to take a classical scattering in the gravity 
dual, characterized by a $b_{max}(s)$, and turn it into a quantum amplitude,
which we can relate to QCD. In section 3 we describe the general formalism 
for calculating $b_{max}(s)$ for black hole formation in the scattering of 
two Aichelburg-Sexl 
(A-S) waves in a background of AdS type. In section 4 we apply this 
formalism for the case of scattering inside AdS and on the IR brane in the 
2-brane RS model. Section 5 is the most important, putting together all the 
pieces of information, and analyzing the various QCD energy regimes, and 
discussing the saturation of the Froissart bound, as well as string corrections
in the gravity dual. In section 6 we conclude and in the Appendix we show 
the details of the calculation of the trapped surfaces being created when 
two A-S shockwaves scatter inside AdS and on the IR brane.

\section{High energy QCD from AdS}

Polchinski and Strassler \cite{ps} have found a simple model for relating
high energy QCD scattering with scattering inside AdS. 

For a conformal field theory, the corresponding
near horizon ($r\rightarrow 0$) metric for the brane configuration is 
\be
ds^2= \frac{r^2}{R^2} d\vec{x}^2 +\frac{R^2}{r^2} dr^2 +R^2 ds_X^2
\ee
The global momentum $p_{\mu}=-i \partial_{\mu}$ 
(momentum for gauge theory scattering, for instance) is thus
related to the local 
inertial momentum (of a local inertial observer in AdS) by
\be
\tilde{p}_{\mu}=\frac{R}{r}p_{\mu} \;\; (\tilde{p}_{\mu}\tilde{p}_{\nu}
\eta^{\mu\nu}=p_{\mu}p_{\nu}g^{\mu\nu})
\ee

Thus high energy is large r and low energy is small r. And then a nonconformal 
gauge theory like QCD 
will just be modified at small r (low energy), at high energy 
remaining conformal. 

Since the low energy cutoff for the conformality of the gauge theory 
should be of the order of the mass of the lightest glueball state $\Lambda$,
the cutoff on the AdS geometry is 
\be
r_{min}\sim R^2 \Lambda
\ee
Thus in theories with mass gap like QCD 
the warp factor becomes bounded, and the simplest
effective description that we are going to use throughout the paper
is just to cut-off the integration over r at $r_{min}$. We can also cut off 
the theory at high energy, and with UV and IR cut-offs we have 
the 2-brane Randall-Sundrum model \cite{rs}. 

Corresponding to the string tension in AdS there is also a gauge theory 
string tension 
$\hat{\alpha}'=(gN)^{-1/2}\Lambda^{-2}$ and 
\be
\sqrt{\alpha '}\tilde{p}=\sqrt{\hat{\alpha}'}p\frac{r_{min}}{r}
\rightarrow \sqrt{\alpha '}\tilde{p}\leq\sqrt{\hat{\alpha}'}p
\label{strtens}
\ee
A glueball correponds in AdS to a state with wavefunction
\be
e^{ipx}\psi(r,\Omega)
\ee
(plane wave in 4d and some wavefunction for r and X). Assuming that 
scattering of gauge invariant states (e.g. glueballs) within Yang-Mills
is equated by AdS-CFT with a scattering inside AdS of the above states;
moreover assuming that the states scatter locally according to the flat 
space amplitude, we get
\be
{\cal A}(p)=\int dr d^5\Omega \sqrt{g}{\cal A}_{string}(\tilde{p}) \prod_i
\psi_i
\label{psrel}
\ee
Then integrating over r corresponds to integrating over scattering energies 
in the local frame ($\tilde{p}$). In this picture $\alpha '$ is a constant, 
as in flat space string theory, but momenta ($\tilde{p}$) ``run'' as r is 
varied, as can be seen from (\ref{strtens}). It is also clear from 
(\ref{strtens}) that one 
can consider $\tilde{p}$ to be fixed and $\alpha '$ to run, but our 
interpretation makes it clear that we are integrating over string theory 
momenta. One should also note that although originally in the formula 
(\ref{psrel}) $A_{string}(\tilde{p})$ was meant as a worldsheet string 
amplitude (with vertex operator insertions), that in flat space gives 
the perturbative spacetime amplitude for scattering particle states 
(e.g. gravitons), we are now extending its meaning. We will first consider 
it at a nonperturbative level, thus we will assume that there is a 
result for the flat space amplitude at the nonperturbative level even if 
we don't know it. Indeed, we are interested in black hole production, for 
which we can calculate the cross section, but not the amplitude, thus in 
the next section we will model the nonperturbative amplitude with a 
simple black disk that reproduces the cross section. Second, the amplitude 
that we will extract has external legs defined only in a flat 4d slice of 
AdS, not in a general direction in AdS, which is however
sufficient for our purposes of convolution with $\psi_i$ according to 
(\ref{psrel}). 

Under the assumption that the local string scattering is 
 dominated by the momenta of the order of the string scale, 
$1/\sqrt{\alpha '}$ (which we will shortly see that it's not so innocent 
as it looks), we get by the above (\ref{strtens}) that
\be
r_{scatt}\sim r_{min} (\sqrt{\hat{\alpha} '} p)
\ee
and if $\sqrt{\hat{\alpha} '} p\gg 1$ (high energy scattering in the 
gauge theory) we see that the integral will be concentrated at $r_{scatt}\gg 
r_{min}$, where 
\be
\psi\sim Cf(r/r_{min})g(\Omega)\sim C(r/r_{min})^{-\Delta}g(\Omega)
\label{wavefct}
\ee
Then
\be
{\cal A}_{QCD}(p)\sim \int dr r^3(\prod_i(\frac{r_{min}}{r})^{\Delta_i})
{\cal A}_{string}
(p R/r)\sim (\frac{\Lambda}{p})^{\sum \Delta_i-4}
\ee
as in QCD, and this result was obtained only
 from conformal invariance. Moreover, for states 
with spin, we replace $\Delta $ with $\tau_i=\Delta_i-\sigma_i$ as in QCD.
So the scaling with momenta comes from the large r asymptotics of the wave 
function, which itself comes from conformal invariance.

But the last scaling relation treats all momenta the same. As we will mostly 
be interested in the small angle regime, $s\gg t$, let's look what happens
for the amplitudes as a function of s and t. 

Given that ${\cal A}_{string}={\cal A}_{string}
(\alpha ' \tilde{s}, \alpha ' \tilde{t})$ 
we take $\nu =\alpha ' |\tilde{t}|=-\alpha ' \tilde{t}$ 
as integration variable ($r=\nu^{-1/2} r_{min} \sqrt{\hat{\alpha} ' t}$) 
and since $s/t=\tilde{s}/\tilde{t}$, we have in the new variable
($\Delta \equiv \sum_i \Delta_i$)
\bea
&&{\cal A}_{QCD}\sim \frac{K}{(\hat{\alpha}' |t|)^{\Delta/2-2}}
\int _0^{\nu_{max}}d\nu \nu ^{\Delta/2-3}{\cal A}_{string}(\nu s/|t|, \nu)
\nonumber\\
&& K= \frac{R^{10}\Lambda^4}{2}(\prod_i C_i) \int \sqrt{g_{(5)}}\prod_i g_i
(\Omega)
\label{ps}
\eea

Note that the main assumption in deriving this was the large r AdS behaviour
of the wavefunction (\ref{wavefct}). We will have to remember this as a 
caveat of the formula, but we will continue nevertheless to use it, it being 
the simplest model we can have of an AdS-QCD relation. 

In this Polchinski-Strassler form for the QCD amplitude, 
 $\nu_{max}=\hat{\alpha '}t$ and the
amplitude to be integrated over is expressed as ${\cal A}(\alpha ' \tilde{s},
\alpha ' \tilde{t})$. 

Let's observe how a few possible behaviours of the string amplitude 
translate into QCD amplitudes, for future use. We will see in the 
next subsection that these types of amplitudes (power laws plus logarithms) 
appear from various possible classical cross sections for scattering in the 
gravity dual, characterized by $b_{max}(s)$. The various $b_{max}(s)$ 
of relevance are analyzed together in section 5.
\bea
&& {\cal A}\simeq A_1 s^{\alpha}t^{\beta}\Rightarrow
{\cal A}_{QCD}\simeq KA_1
 s^{\alpha}t^{\beta}(\frac{\hat{\alpha}'}{\alpha '})^{\alpha 
+\beta}\frac{1}{\Delta /2-2+\alpha +\beta}
\nonumber\\
&&{\cal A}\simeq A_2 s^{\alpha }t^{\beta} log (\alpha ' t) \Rightarrow 
\nonumber\\&&
{\cal A}_{QCD} \simeq K A_2 
 s^{\alpha}t^{\beta}(\frac{\hat{\alpha}'}{\alpha '})^{\alpha +\beta }
\frac{1}{\Delta /2-2+\alpha +\beta}
(log (\hat{\alpha}'t)-\frac{1}{\Delta /2-2+\alpha+\beta})\nonumber\\
&& {\cal A}\simeq A_3 log (\alpha ' s) t^{\beta} \Rightarrow {\cal A}_{QCD}
\simeq K A_3
\frac{t^{\beta}}{\Delta/2-2+\beta}(\frac{\hat{\alpha}'}{\alpha '})^{\beta}
(log(\hat{\alpha}'s)-\frac{1}{\Delta /2-2+\beta})
\label{behav}
\eea

Giddings \cite{gid} 
points out that as one increases the energy of the gauge theory 
scattering, by (\ref{strtens}) one increases also the relevant energy 
in string theory. In (\ref{strtens}), we have seen that corresponding to 
the string scale $1/\sqrt{\alpha '}$ there is a gauge theory scale 
$1/\sqrt{\hat{\alpha}'}$.
But there are three further (higher) energy scales (in the 
case when the string coupling $g_s$ is small but $g_sN$ is large).

The first is the Planck scale
\be
M_P\sim g_s^{-1/4} (1/\sqrt{\alpha '})=\frac{N^{1/4}}{R}=N^{1/4}\sqrt{\frac{
\Lambda}{r_{min}}}
\ee
Note that we can rescale 4d coordinates such that $r_{min}=R$, which, since 
$\hat{\alpha}'=\Lambda^{-2}(g_sN)^{-1/2}$, translates also into 
$\hat{\alpha}'=\alpha '$. Then $M_P= N^{1/4}\Lambda$. 

In any case, the Planck scale corresponds in 
gauge theory to 
\be
\hat{M}_P=g_s^{-1/4}/\sqrt{\hat{\alpha}'}=N^{1/4}\Lambda
\ee
which is the scale at which (real) black holes start to form. 

Giddings \cite{gid} proposes that afterwards the black hole production 
cross section 
is approximated by $\sigma\simeq \pi r_H^2\sim E^{2/(d-3)}$ ($E^{2/7}$ if
we have approximately 10d flat space). As we can see, this is based on the 
simple geometrical picture of a static black hole at a given point in AdS, 
with the cross section equaling its horizon area. We will try to see whether 
this picture is valid. 

The second scale is the string correspondence principle cross-over scale, at 
which one has to stop talking about string intermediate states and 
instead use black hole 
virtual states. That scale is
\be
E_c\sim g_s^{-2}/\sqrt{\alpha '}=g_s^{-7/4}M_P= N^{7/4}M_P/(g_sN)^{7/4}=
N^2\sqrt{\frac{\Lambda}{r_{min}}}\frac{1}{(g_sN)^{7/4}}
\ee
or in the gauge theory 
\be
\hat{E}_c\sim \frac{N^2\Lambda}{(g_sN)^{7/4}}
\ee

The third energy regime is the most interesting, attained when the 
size of the black hole $r_H$ that was created reaches the AdS size R:
\be
E\sim M_P^8 r_H^7\;\; (M_P^{d-2}r_H^{d-3})\rightarrow E_R=M_P^8R^7
\label{energy}
\ee
or in the gauge theory,
\be
\hat{E}_R= N^2 \Lambda
\ee
At that energy, the behaviour of $r_H$ with E changes from (\ref{energy}) 
and then so does the cross-section. The proposal of \cite{gid} is that 
after this scale we have the onset of the Froissart bound in the gauge theory.

Let us now note that at least after the onset of the Froissart behaviour 
we have black holes being created with Schwarzschild radius greater than 
the size of AdS, so one might ask how come we are still using  the 
Polchinski-Strassler formula (\ref{psrel}) which implies some locality 
for the scattering in the extra dimensions? In fact, in this regime the 
PS formula becomes less and less relevant, since the effective scattering 
region (where the PS integral is concentrated) is small and close to the 
IR cut-off (as we will see in the following), 
and the size of the black hole is larger than it. So in this limit, 
the scattering inside AdS becomes more and more classical (the fluctuations 
due to the PS integral become less than the size of the black hole), and the 
classical cross section calculations in this paper become more 
directly relevant.

\subsection{Black disk calculations}

In the following calculations, we will analyze
 a classical scattering with black hole 
creation, out of which we get a classical value for a $b_{max}(s)$, 
and correspondingly a cross section for black hole creation, $\sigma =
\pi b_{max}^2$. But in order to use the Polchinski-Strassler formalism
and relate it to QCD, 
we must find a quantum amplitude generating the same AdS cross section. 
So we will study first how we get a quantum amplitude out of the classical 
picture. 

Ideally, one should do a quantum calculation for the amplitude of the AdS 
scattering, not just for the cross section (for which we were able to use the 
formalism of AS shockwave scattering described in the next section)
but that would require knowledge of the full nonperturbative 
quantum gravity. As we don't have that, we will try to make a model 
for the amplitude that reproduces the classical cross section calculation,
$\sigma=\pi b_{max}^2(s)$. 

The simplest thing one can do is to create an eikonal that corresponds to a 
black disk, that is 
\be
Re(\delta(b,s) )=0\;\;\; Im (\delta(b,s) )=0,\;b>b_{max}(s);\;\;\;
Im (\delta(b,s) )=\infty , \; b<b_{max}
\ee

Then the eikonal amplitude at a fixed flat 4d slice 
inside AdS that reproduces the classical $b_{max}(s)$ for scattering 
constrained to lie in the slice  in s,t variables is  
\bea
\frac{1}{s} {\cal A}(s,t) &=&-i \int d^2 b e^{i\vec{q}\vec{b}} (e^{i\delta}-1)
=i\int_0^{b_{max}(s)}bdb \int _0^{2\pi} d\theta e^{iqb \cos\theta}
\nonumber\\
&=& 2\pi i\frac{b_{max}(s)}{\sqrt{t}} J_1(\sqrt{t}b_{max}(s))
\label{eikonal}
\eea
We will take the amplitude (\ref{eikonal}) to be the amplitude for 
scattering inside AdS space, parallel to a fixed 4d slice. Here $b_{max}(s)
$ will be determined from black hole production in the scattering of two 
Aichelburg-Sexl shockwaves inside the curved AdS background. 

In general, the notion of scattering inside AdS space and the associated 
notions of S matrix and in and out states are hard (if not impossible) to 
define. In particular, the original definition of the AdS-CFT correspondence 
states that a set of good quantities that can be related to the boundary CFT
are the corellators with legs on the AdS boundary, not S matrices. There were 
several attempts to define S matrices in AdS-CFT by taking the flat space 
limit of AdS, e.g. \cite{polc,suss,gidd}, but none was entirely satisfactory.
What we are proposing here is less radical. First of all, unlike these other 
cases, we are not dealing with global AdS, and we are not relating in and out 
states defined at the boundary of AdS (which would be in the Poincare patch 
at $r=0$ and $r=\infty$). Rather, we are looking at scattering 
amplitudes  that happen in a 4d slice, at fixed r. Second, when we talk 
about cross sections we refer to 4d quantities, inside the 4d slice. 
We are not attempting to define a 5d cross section for scattering inside 
AdS. Here the 5th coordinate of AdS, r, has just the usual role of energy 
scale (according to the usual UV-IR relation in AdS-CFT), as can be easily 
seen in (\ref{psrel}).

We note that if $\sqrt{t} b_{max}(s)\ll 1$, the result becomes 
$\pi b_{max}^2$, so if we take the imaginary part of the forward 
(t=0) scattering amplitude we get
\be
\frac{1}{s}Im{\cal A}_{elastic}(k_1,k_2\rightarrow k_1, k_2)= \sigma_{tot}
(k_1,k_2\rightarrow anything)= \pi b_{max}^2
\ee
But we still need to integrate the amplitude over the AdS slice, using the 
P-S formula. 

We use the Polchinski-Strassler formula (\ref{ps}), i.e.
\be
{\cal A}_{QCD}=\frac{K}{(\hat{\alpha}'t)^{\Delta/2-2}}\int_0^{\nu_{max}=\hat{
\alpha}'t}d\nu \nu^{\Delta/2-3}{\cal A}(\nu \frac{s}{|t|},\nu)
\nonumber
\ee
Upon inserting the eikonal amplitude (\ref{eikonal}) into (\ref{ps}), we get 
\be
{\cal A}_{QCD}= \frac{s}{|t|}\frac{\sqrt{\alpha '}2\pi i K}
{(\hat{\alpha}'t)^{\Delta/2-2}}
\int_0^{\nu_{max}=\hat{\alpha}'t}d\nu \nu^{\frac{\Delta -5}{2}}
b_{max}( \frac{\nu s}{\alpha '|t|})J_1(\sqrt{\frac{\nu}{\alpha '}}b_{max}
(\frac{\nu s}{\alpha '|t|}))
\ee
Let us now look particular cases that will be of interest later on.
In all the three cases we will study, the result for ${\cal A}_{QCD}$
will be of the type {\em mass dimension -2 constant} ($K a_1$, $K a_2^{2/(1+
2\beta)}$, 
$K a_3^2$) $\times$ {\em function of} ($\hat{\alpha '} s, \hat{\alpha
'} t, \Delta$ and c). Here c stands for 
a dimensionless number calculable in the gravity dual
($a_1/\alpha'$, $a_2^{2/(1+2\beta)}/\alpha '$, $a_3^2/\alpha'$). 
Here $a_1, a_2, a_3$ are calculable in the gravity 
dual and K encodes the details of the transition to QCD, i.e it
 depends heavily on
the details of the IR cut-off of the gravity dual, as can be seen from its 
expression in (\ref{ps}).

{\em Approximately 4d case}: $b_{max}(s)=a_1\sqrt{s}$. Using 
\be
I(a,b) 
= \int_0^1x^b J_1(a x)dx = \frac{a \;\; {}_1F_2(1+b/2; 2,2+b/2; -a^2/4)}{4+2b}
\ee
we get
\be
{\cal A}_{QCD}=-\frac{2\pi i a_1^2 K}{\Delta }\frac{\hat{\alpha}'{}^2s^2}
{\alpha '}
{}_1F_2 (\frac{\Delta }{4};2,\frac{\Delta+4}{4}, -\frac{a_1^2\hat{\alpha} '^2st
}{4{\alpha '}^2})=K a_1 \times fct.(\frac{a_1}{\alpha '}, \hat{\alpha '} s,
\hat{\alpha '}t, \Delta)
\ee
To evaluate the hypergeometric function at large argument (variable), we use 
that 
\be
J_1(ax)\simeq - \sqrt{\frac{2}{\pi a x}}\cos (ax +\pi/4)
\ee
and then 
\bea
I(a,b)&\simeq &\frac{1}{\sqrt{\pi a}}\frac{1}{2b+1}[{}_1F_1(b+1/2; b+3/2; ia)
(1+i)- {}_1F_1(b+1/2;b+3/2; -i a )(1-i)]\nonumber\\
&\simeq &\frac{2\sqrt{2}}{\sqrt{\pi a}}\frac{\Gamma(b+3/2)}{2b+1}
\cos (b\pi/2) a^{-(b+1/2)}
\eea
We get
\bea
{\cal A}_{QCD}&\simeq &i k \hat{\alpha} ' t(\frac{\hat{\alpha}'t}{\alpha '})
^{-\frac{\Delta -2}{2}}(\frac{s}{t})
^{\frac{8-\Delta}{4}}= i k (\frac{\hat{\alpha}'}{\alpha '} )
^{-\frac{\Delta-2}{2}}s^{\frac{8-
\Delta}{4}}t^{-\frac{\Delta}{4}}\hat{\alpha} '
\nonumber\\
k&=& 2\sqrt{2\pi }\Gamma (\frac{\Delta-3}{2})a_1^{\frac{4-\Delta}{2}}
\cos (\frac{\pi (\Delta-4)}{4})
\eea
in the limit $\frac{a_1^2}{{\alpha '}^2}
\hat{\alpha }'^2 st\gg 1$. In the opposite limit, 
$\frac{a_1^2}{{\alpha '}^2}\hat{\alpha}'^2 st \ll 1$, we get 
\be
{\cal A}_{QCD}\simeq -\frac{2\pi i a_1^2K}{\Delta }\frac{\hat{\alpha}'{}^2s^2}
{\alpha '}
\ee

{\em Flat d dimensional case}: $b_{max}(s)=a_2 s^{\frac{1}{2(d-3)}}= 
a_2 s^{\beta}$

Using the same formulas, we get 
\bea
&&
{\cal A}_{QCD}= \frac{2\pi i a_2^2 K}{4\beta +\Delta-2} (\frac{\hat{\alpha}'s}
{\alpha '})^
{2\beta} \hat{\alpha }' s {}_1F_2(1+\frac{\Delta-4}{2(2\beta+1)};2,2+
\frac{\Delta-4}{2(2\beta+1)}; -\frac{a_2^2}{4}\frac{\hat{\alpha}'t}{\alpha '}
(\frac{\hat{\alpha}'s}{\alpha '})^{2\beta})\nonumber\\&&
= K a_2^{\frac{2}{1+2\beta}} \times fct.(\frac{a^{\frac{2}{1+2\beta}}}{\alpha '
}, \hat{\alpha '}s, \hat{\alpha ' t}, \Delta)
\eea
It has the limit
\bea
{\cal A}_{QCD} &\simeq &
 \bar{K} (\frac{\hat{\alpha}' s}{\alpha '}) ^{-\frac{\beta(\Delta-4)}
{2\beta+1}} (\frac{\hat{\alpha }'t}{\alpha '})^{\frac{2-\Delta 
-4\beta }{2(2\beta+1)}}\hat{\alpha}' s \nonumber\\
\bar{K}&=& iK\frac{2\sqrt{2\pi }}{\beta +1/2}\frac{\Gamma (\frac{\Delta-4}
{2\beta +1}+\frac{1}{2})}{a_2^{\frac{\Delta-4}{2\beta+1}}}
\cos [\frac{\pi}{2}(\frac{\Delta-4}{2\beta +1})], \;\;\;
a_2^2 \frac{\hat{\alpha}'t}{\alpha '} (\frac{\hat{\alpha}'s}{\alpha '})
^{2\beta}\gg 1
\eea
If $d=5$, so $\beta=1/4$, we get 
\be
{\cal A}_{QCD}\sim k' s^{-\frac{ (\Delta-4)}{6}} t^{\frac{1-\Delta}{3}}
\hat{\alpha}' s 
\ee
Another limit is 
\be
{\cal A}_{QCD}\simeq \frac{2\pi i a_2^2K}{4\beta +\Delta-2}(\frac{
\hat{\alpha}'s}{\alpha '})^
{2\beta} \hat{\alpha '}s ,\;\;\;
\frac{\hat{\alpha}'t}{\alpha '} (\frac{\hat{\alpha}'s}{\alpha '})
^{2\beta} \ll 1
\ee

{\em Log behaviour: } $b_{max}(s)=a_3 \;\;log (s)$, that 
would correspond to the onset of the Froissart bound, at least in the 
calculation in \cite{gid}. 
We get 
\bea
&&{\cal A}_{QCD}= \sqrt{\alpha '}(\hat{\alpha}' s)
\frac{2\pi i a_3 K}{\sqrt{\hat{\alpha}'t}}\int_0^1
dy y^{\frac{\Delta -5}{2}}[ln y +A]J_1[B\sqrt{y}(ln y +A)]\nonumber\\
&&= K a_3^2 \times fct. (\frac{a_3^2}{\alpha '}, \hat{\alpha '} s, \hat{\alpha
' } t, \Delta)\nonumber\\
&& A= ln (\frac{\hat{\alpha}'s}{\alpha '}),\;\;\; B=a_3\sqrt{
\frac{\hat{\alpha }'t}{\alpha '}}
\eea
that unfortunately cannot be algebraically solved exactly. 

But in the case of $A$ large, $c=AB$ large, we can do the integral, using the 
large argument expansion of the Bessel function, and using 
\be
\int_0^1 dx x^b ln x (\cos cx -\sin cx)\simeq -\frac{1}{c}\int_0^1 dx x^{b-1}
(\sin cx -\cos cx)\simeq -\frac{\Gamma(b)\sqrt{2}}{c^{b+1}}\cos (\frac{\pi b}{2
}-\frac{\pi}{4})
\ee
and we get
\be
{\cal A}_{QCD}\simeq 4\sqrt{2\pi}i K \Gamma (\Delta -9/2)
\frac{(a_3\sqrt{\frac{\hat{\alpha}'t}{\alpha '}}ln (\frac{\hat{\alpha}'s}
{\alpha '}))^{4-\Delta}}
{\hat{\alpha}'t/\alpha '}[(\Delta
-9/2)\cos (\frac{\pi}{2}(\Delta-3))+2\frac{\cos(\frac{\pi}{2}(\Delta-5))}{
ln (\frac{\hat{\alpha}'s}{\alpha '})}]
\ee

Also for $A\gg 1$, but $c=AB \ll 1$, we can expand the Bessel function 
for small argument and do the integration to obtain
\be
{\cal A}_{QCD}\simeq \hat{\alpha }' s 2\pi i a_3^2K
[\frac{(ln (\frac{\hat{\alpha}'s}{\alpha '}))^2 }
{\Delta-2}-\frac{2}{(\Delta -2)^2}ln (\frac{\hat{\alpha}'s}{\alpha '})]
\ee

Note that for all 3 cases, when we have used the small argument expansion
of the Bessel function, we have integrated $\pi b_{max}(s)^2$ with the PS 
formula. Correspondingly, in the last case we studied, we obtain 
$\sigma \sim a_3^2 (ln (s))^2$, as expected. 

We should also note that within this section we have used the simplest model 
of black disk for turning the classical scattering into a quantum scattering 
amplitude, but for the last behaviour (the Froissart bound) there are 
other choices. For instance, in \cite{km}, there are given a few forms for 
the eikonal which when integrated give the Froissart behaviour for the 
cross section. One of them is of possible relevance here. The eikonal 
\be
Im \; \delta (b,s)=\frac{\pi \lambda}{B(s)}s^{\bar{\Delta}} exp \{ -\frac{b^2}{
B(s)} \}
\label{nte}
\ee
contains a parameter analogous to the classical maximum impact parameter
$b_{max}(s)$ we have used:
\be
B(s)= 2 a\; ln \; s +c
\ee
where however $a$ and $d=\bar{\Delta} +1$ are related to the Pomeron trajectory
$\alpha_P(t)=d+ a\; t$. Integrating this eikonal also gives the Froissart 
behaviour for the total cross section
\be
\sigma _t(s) \simeq 4\pi \;a \; \bar{\Delta} ln^2 s+ o(ln \; s)
\ee 
We will restrict ourselves to the simplest black disk eikonal
as we did in this section, as there is 
no physical argument on why to choose a particular {\em nontrivial} eikonal
(in the full nonperturbative quantum gravity one should be able to calculate
the actual nontrivial eikonal, of course, but in practice there seems to be 
no reason to prefer a nontrivial phenomenological eikonal
like the one in (\ref{nte}) over another).

\section{Black hole creation in high energy scattering; general formalism}

As we saw in the previous section, at sufficiently high energies in the 
gauge theory, in the gravity dual we will have an inelastic scattering 
with black hole creation. So we have to be able to analyze the black hole 
creation in a general background.  

In \cite{kn} we have extended the formalism in \cite{eg} for calculating 
black hole creation cross sections via analyzing the scattering of two 
Aichelburg-Sexl shockwaves inside a curved background of AdS type. 
Here we review the formalism in order to apply it in the next section.

The Aichelburg-Sexl shockwave \cite{as}
\be
ds^2 =-dx^+dx^-+(dx^+)^2\delta(x^+) \Phi(x^i) +\sum_{i=1}^{d-2} (dx_i)^2
\ee
has as a source a massless particle of momentum p (``photon''), with 
\be
T_{++}= p \delta ^{d-2}(x^i)\delta (x^+)
\ee
In flat space, the Einstein equation $R_{++}= 8\pi G T_{++}$ implies
\be
 \partial _i^2 \Phi (x^i)=-16\pi G p\delta^{d-2}(x^i)
\ee
($\Phi$ is harmonic with source).

't Hooft \cite{thooft,dh} has argued that one can describe the scattering 
of two massless particles at energies close to (but under) the 
Planck scale, ($m_{1,2}\ll M_P, Gs\sim 1$, yet $Gs<1$) as follows.
Particle two creates a 
massless shockwave of momentum $p_{\mu}^{(2)}$ and particle one follows a 
massless geodesic in that metric. He has shown that the S matrix 
corresponds to a gravitational Rutherford scattering (single graviton 
being exchanged).

At higher energies ($Gs\gg 1$), one has to consider that both massless 
particles create A-S shockwaves, and this nonlinear process is hard to 
compute. At most one can compute the metric perturbation away from the
interaction point as in \cite{dp}.
But one can use a formalism to give a lower bound on 
the size of the black hole being created, and estimate the maximum impact 
parameter that forms a black hole. 

We will be interested in scattering that occurs in a gravitational 
background of AdS type, maybe with a string-corrected source. So we will 
use a general $\Phi$ and general dimensionality, and a background of the type 
(the notation is for the one brane RS model, but it is easy to generalize 
for the AdS and 2-brane RS model cases)
\be
ds^2=e^{-2|y|/l}[-dudv +dx_i^2]+dy^2
\label{UVmetric}
\ee
Let us denote $e^{-2|y|/l}=A$ and let $g_{ij}=A\bar{g}_{ij}$ for the 
metric in both x and y coordinates (transverse). 

We will analyze the collision of two A-S waves in this background, one 
moving in the $x^+$ (u), one moving in the $x^-$ (v) direction. The metric 
in the collision region cannot be calculated exactly even in flat space, 
but there is an alternative for checking for the presence of a black hole 
in the future of the collision.

Due to a suggestion made originally by Penrose, one can find a trapped 
surface at the interaction point u=v=0, that is, a closed D-2 dimensional 
surface the outer normals of which (in both future-oriented directions) 
have zero convergence. By a GR theorem, we know that there will be a horizon 
forming outside the trapped surface, therefore of area at least as big as the 
trapped surface area. 

The metric with a single A-S wave moving in the $\bar{v}$
 direction (at $u=0$) in the given background is (see \cite{kn}
for more details)
\be
ds^2=e^{-2|\bar{y}|/l}(-d\bar{u}d\bar{v}+d\vec{x}^2+\Phi(\vec{x}, y)
 \delta(\bar{u})
d\bar{u}^2)+d\bar{y}^2
\ee
It is useful to perform a transformation of coordinates to eliminate
the delta function singularity in the metric. One finds 
the transformation
\bea
&& \bar{u}= u\nonumber\\
&&\bar{v}= v + \Phi\theta (u) +\frac{u\theta(u)}{4} (\partial_i \Phi
 \partial_j \Phi  
\bar{g}^{ij}+A (\partial_y \Phi)^2)\nonumber\\
&& \bar{x}^i = x^i +\frac{u\theta(u)}{2}\bar{g}^{ij}\partial_j \Phi \nonumber\\
&&\bar{y}=y +\frac{u\theta(u)}{2}A\partial_y \Phi
\eea
giving 
\bea
&&ds^2=A[-dudv +dx_i^2+u\theta(u)\partial_i\partial_j \Phi dx^i dx^j]
\nonumber\\&&+dy^2[1+
u\theta(u)A\partial_y^2\Phi ]+dydx^i u\theta (u) A \partial_i\partial_y \Phi +
dydA u\theta (u) \partial_y \Phi +o(u^2)
\eea
where
\be
A= e^{-2|\bar{y}|/l}+o(u)^2\Rightarrow 
A|_{u=0}=e^{-2|y|/l};\;\;\;dA|_{u=0}=-\frac{2}{l}A[dy+\frac{A}{2} \partial_y  
\Phi du]
\ee

When taking two A-S waves, one in $\bar{u}$ and one in $\bar{v}$, the metric
for $\bar{u}\leq 0, \bar{v}\leq 0$ (before the collision) contains the 
linear superposition of the 2 waves, i.e. $\Phi_1 \delta (\bar{u})d\bar{u}^2
+\Phi_2\delta(\bar{v})d \bar{v}^2$, and a trapped surface will form at $\bar{u}
=\bar{v}=0$.
The trapped surface in the new coordinates (u, v) is composed of two ``disks''
$S_1$ and $S_2$ glued 
on to their common boundary C, namely 
 
``disk'' 1- $\{ v=-\Psi_1(\vec{x}), \;\; u=0\}$, ($\Psi_1=0$ on C)

``disk'' 2- $\{ u=-\Psi_2(\vec{x}),\;\; v=0 \}$  ($\Psi_2=0$ on C).

The null geodesics through the first disk,
$\{v=-\Psi(\vec{x}), \;\; u=0 \}$ are 
defined by the tangent vector
\be
\xi =\dot{u}\frac{\partial}{\partial u}+\dot{v}\frac{\partial}{\partial v}
+\dot{x}^i\partial _i
\ee
One finds that (by calculating $\dot{u}, \dot{v}, \dot{x}$ and then lowering
indices)
\be
\xi=-\frac{1}{4}\bar{g}^{ij}\partial_i\Psi\partial_j\Psi du 
-dv -\partial_i \Psi dx^i
\ee
and therefore the convergence of the null normals 
defined by $\xi$, $\theta=g^{ij}D_i\xi_j$ is 
(see \cite{kn} for more details)
\be
\theta=-\nabla^2(\Psi_1-\Phi_1)
\ee
and similarly for the second surface, and 
\be
\nabla^2=\frac{1}{A}\nabla^2_x+\partial_y^2-\frac{d}{l}sgn(y)\partial_y
\ee
Here $\Phi_1$, $\Phi_2$ are the profiles of the two waves, whose centers 
are separated by the impact parameter $\vec{b}=\vec{x}_1-\vec{x}_2$.

At b=0 and in flat D dimensions ($l \rightarrow \infty$)
one can choose $\Psi_i=\Phi_i$, and then as we can easily see, 
the trapped surface $S=S_1 U S_2$ corresponds as advocated to 
the interaction point $\bar{u}=
\bar{v}=0$. 

The common boundary C is defined then by $\Psi= \Phi=c$ (constant) or rather,
one redefines $\Psi\rightarrow \Psi-c$ to have $\Psi=0$ on the boundary.
We will keep the constant for convenience, but remember that we should 
subtract it at the end. 
The continuity condition for the normal geodesics $\xi$ along C gives then
\be
\bar{g}^{ij}\partial_i\Psi\partial_j\Psi =4 \Rightarrow
(\nabla \Psi)^2+A(\partial_y\Psi)^2=4
\ee

At b=0 and in flat D dimensions, the $\Psi=\Phi= c$ and the continuity 
condition are compatible with the boundary C being a circle (r=const), and 
then the continuity condition fixes the radius of the circle. 

In curved background  we can't choose $\Psi=\Phi$ anymore,
and the shape of C is fixed by requiring that the two equations are compatible.

Therefore we write for the trapped surface (surface of zero convergence,
$\theta=0$)
\be
\Psi=\Phi+\zeta \Rightarrow \nabla^2\zeta=0
\ee

Now the trapped surface  $f(r, y)=0$ is defined by both 
$\Psi=C$ (const) and by $\bar{g}^{ij}\partial_i\Psi\partial_j\Psi =4 $.

We will find the shape of the surface 
 perturbatively in y, away from the flat 4 dimensions at y=0. 
Expand $\zeta $ near y=0 as 
\be
\zeta=\zeta_0(r)+\zeta_1(r)y+\frac{y^2}{2}\zeta_2(r)
\ee
Then at y=0 $\nabla^2\zeta=0$ implies
\be
\partial_x^2 \zeta_0(r)+\zeta_2(r)-\frac{d}{l}\zeta_1(r)=0
\ee
and we don't want to upset the flat space solution, so we will take 
$\zeta_0=0$.
Then
\be
\Psi=f+a y +\frac{y^2}{2}g+...
\ee
where $f=\Phi_{|y=0}$, $a=\partial_y \Phi|_{y=0}+ \zeta_1$ and 
$g=\partial_y^2\Phi|_{y=0}+\frac{d}{l}\zeta_1$.

And one has to match the equations 
\be
\Psi=C= f+ a y +\frac{y^2}{2}g+...
\label{trapped}
\ee
and 
\be
4= (f'+y a'+\frac{y^2}{2}g'+...)^2+(1-\frac{2y}{l}+2\frac{y^2}{l^2}+...)
(a+yg+...)^2
\ee

At nonzero impact parameter b, there are two distinct $\Phi_1$ and $\Phi_2$ 
(with 
different centers), and a single curve C, so we can't take $\Psi_i= \Phi_i$
even in flat D dimensions. The problem of finding C together with the functions
$\Psi_i$ is complicated, but we have found instead an approximation scheme. 
The continuity condition is now $\nabla \Psi_1\cdot \nabla 
\Psi_2=4$, but we approximate 
the size of C by putting $\Psi_i=\Psi(\vec{x}-\vec{x}_i)$, with $\Psi$ being 
the b=0 value. This gives a continuity equation 
\be
4= (1-\frac{b^2}{2\rho_c^2})
[(f'+y a'+\frac{y^2}{2}g'+...)^2+(1-\frac{2y}{l}+2\frac{y^2}{l^2}+...)
(a+yg+...)^2]|_{r=\rho_c} 
\ee

Now one can say that the area of the real trapped surface satisfies
\be
S\ge b\sqrt{\rho_c^2-\frac{b^2}{4}}
\ee
and more importantly we can estimate a maximum b for which  a trapped 
surface forms.

\section{ Black hole creation inside AdS and on the RS IR brane}

In this section we will apply the formalism of the previous section to the 
case of A-S shockwaves inside AdS and on the IR brane in the 2-brane RS 
model. The details of the calculation of the trapped surface are found in 
the Appendix, so here we will only show the general features.

Note that the application of the general formalism for scattering on the 
UV brane of the RS model (as well as for large extra dimensions), and thus for 
possible applications to accelerators was done in \cite{kn}. Here we 
concentrate on the cases relevant for QCD, namely (as we will discuss 
in detail in the next section) scattering inside AdS and on the IR 
brane.  

We have already derived the form of the A-S wave inside AdS in \cite{kn}.
The function $\Phi$ for an A-S wave centered around $r=0, y=y_0$ is
\bea
\Phi(r,y)&=& \frac{8G_{d+1}l}{(2\pi)^{\frac{d-4}{2}}}p 
\frac{e^{\frac{dy}{2l}}e^{\frac{4-d}{2l}y_0}
}{r^{\frac{d-4}{2}}}\int_0^{\infty}dq 
q^{\frac{d-2}{2}}J_{\frac{d-4}{2}}(qr) K_{d/2}(e^{y/l}lq)I_{d/2}(e^{y_0/l}lq)
\;\;\;y>y_0\nonumber\\
&=&\frac{8G_{d+1}l}{(2\pi)^{\frac{d-4}{2}}}p 
\frac{e^{\frac{dy}{2l}}e^{\frac{4-d}{2l}y_0}
}{r^{\frac{d-4}{2}}}\int_0^{\infty}dq 
q^{\frac{d-2}{2}}J_{\frac{d-4}{2}}(qr) I_{d/2}(e^{y/l}lq)K_{d/2}(e^{y_0/l}lq)
\;\;\;y<y_0\nonumber\\&&
\eea

Let us therefore 
first derive the A-S wave solution inside the 2-brane RS metric
(the one used by Giddings in his calculation of the Froissart bound),
with the A-S situated on the IR brane. It will
be different from the Emparan \cite{emp}
solution for the A-S wave on the brane in the 1 brane RS. 
We will use the same formalism used for the 1-brane RS case by Emparan 
and for AdS by \cite{kn}. We would still use the AdS metric
\be
ds^2=e^{-2y/l}d\vec{x}^2+dy^2
\ee
as in \cite{gid}, but then $y\in (-\infty, 0)$ and
 the IR brane is located at y=0 and the UV brane at $y=-|y_{UV}|$. The metric 
would be valid only in $(-|y_{UV}|, 0)$. For a general metric satisfying 
the required boundary conditions we will take
\be
ds^2=e^{2|y|/l}d\vec{x}^2+dy^2
\label{IRmetric}
\ee
It matches with the above for its domain of validity (y negative).

In complete analogy with the calculation in \cite{emp} and \cite{kn}, 
we obtain an equation for $h(q,y)= \int d^{d-2}\vec{x} e^{iq\cdot x}\Phi(x,y)$ 
which is
\be
h(q,y)''+\frac{d}{l}sgn(y) h(q,y)'-q^2 e^{-2|y|/l}h(q,y)=0
\ee
with solution 
\be
Ae^{-\frac{d|y|}{2l}}(I_{d/2} {\rm or}K_{d/2})(e^{-|y|/l}lq)
\ee
and imposing normalizable behaviour at $y=\pm \infty$ (at the UV brane, 
if that is moved to infinity) we restrict to $I_{d/2}$. Then imposing the 
jump condition at y=0 (IR brane), that is putting 
the source of the wave on the 
IR brane as in \cite{gid}, we finally get 
\bea
\Phi (r,y)
&=& \frac{4G_{d+1}p}{2\pi}
e^{-\frac{d|y|}{2l}}\int \frac{d^{d-2}\vec{q}}{(2\pi)^{d-2}}e^{i\vec{q}
\vec{x}}\frac{I_{d/2}(e^{-|y|/l}lq)}{qI_{d/2-1}(lq)}\nonumber\\
&=&\frac{4G_{d+1} p}
{(2\pi)^{\frac{d-4}{2}}}\frac{e^{-\frac{d|y|}{2l}}}{r^{\frac{d-4
}{2}}}\int_0^{\infty} dq q^{\frac{d-4}{2}}J_{\frac{d-4}{2}}(qr)\frac{I_{d/2}
(e^{-|y|/l}lq)}{I_{d/2-1}(lq)}
\eea
(in d=4, $4G_{d+1}p= R_s l$).
By comparison, Giddings has an $h_{00}$ (Newton potential) obtained also 
from a wave equation with sources on the IR brane \cite{gid}, 
except his source was a 
static mass (black hole), whereas for us it is a photon
\be
h_{00}\sim e^{\frac{dy}{2l}}\int \frac{d^{d-1}\vec{p}}{(2\pi)^{d-1}}
e^{i\vec{p}\vec{x}}\frac{J_{d/2}(ip l e^{y/l})}{ipJ_{d/2-1}(ipl)}
\ee
and since $J_{\nu}(ix)=I_{\nu}(x)$ and y is negative, we have almost the 
same solution, except for us the integration is only over d-2 transverse 
coordinates (as for a massless particle), whereas for Giddings the integration 
is over d-1 transverse coordinates, as appropriate for a massive particle.

Let us then analyze the AdS scattering. We are interested
in the large r behaviour ($e^{y/l}l/r\ll 1$ and $y=y_0$) which is relevant
for the Froissart bound that we are after. In that case, 
\be
\Phi= \frac{\bar{C}l^4}{r^6} e^{\frac{2}{l}(2y+y_0)}
\ee
We find (see the Appendix for details; here $\epsilon\equiv y-y_0$)
\bea
&& f=\Phi|_{\epsilon=0}= \bar{C} \frac{l^4}{r^6}e^{6y_0/l}\nonumber\\
&& a=\frac{4}{l}f+\zeta_1\nonumber\\
&& g=\partial^2_{\epsilon}\Phi|_{\epsilon=0}+\frac{4}{l}\zeta_1
=\frac{16 f}{l^2}+ \frac{4}{l}\zeta_1\nonumber\\&&
\Rightarrow a= -\frac{\alpha}{l} f=-\frac{16}{l} f
\eea
As we have explained in the Appendix, we actually get two solutions 
for the trapped surface. There is one solution for which $a$ is neglijible,
thus that solution corresponds to what we would have if the scattering 
was four dimensional (and just $\Phi$ was obtained from the 5d equations).
But in the $r \gg l $ limit, the AdS warping is very large, and energy 
scales become larger away from the 4d slice, thus we expect the size of the 
black hole being created (and thus of the trapped surface) is increased.
And so we will take the solution that takes into account the 5d scattering,
solution that implies a larger horizon. 

The same situation will be encountered in the case of scattering on the IR 
brane. The space will then be very non-4-dimensional in the $r\gg l$ limit
(in a sense, it will be anti-4-dimensional), thus we expect that the 
size of the trapped surface will be increased also with respect to 
pure 4d scattering. This situation is to be contrasted with the scattering 
on the UV brane we analyzed in \cite{kn}, in which case for $r\gg l$ the 
space was approximately 4 dimensional (the warping is going down away from 
the brane), and correspondingly we found a solution which had just 
small corrections to the four dimensional scattering. 

The continuity condition for the larger trapped surface is thus 
\be
\frac{|\alpha| f}{2l}e^{-y_0/l}=1
\ee
so that 
\be
r=r_{max}=l e^{y_0/l}[\frac{3 R_s}{le^{y_0/l}}]^{1/6}
\ee
and since in 4 dimensions
$R_s=2G_4 \sqrt{s}$, we have that $r_{max}\sim s^{1/12}$.

If we introduce a nonzero impact parameter the continuity condition 
becomes 
\be
(\frac{3 R_s l^5 e^{5y_0/l}}{r^6})^2 (1-\frac{b^2}{2r^2})=1
\ee
and with $r^2=x, (3 R_s l^5 e^{5y_0/l})^2=a$ we get the equation 
\be
x^7-ax +a\frac{b^2}{2}=0
\ee
The maximum b for which it has a solution is (using $f(x_0)=0$ for $f'(x_0)=0$)
\be
b_{max}^2= \frac{12}{7}[\frac{a}{7}]^{1/6}
\Rightarrow 
b_{max}=7^{-1/12}\sqrt{\frac{12}{7}}l e^{y_0/l}[\frac{3 R_s}{le^{y_0/l}}]^{1/6}
=\frac{1}{7^{1/12}}\sqrt{\frac{12}{7}}r_{max}
\ee
so that $b_{max}\sim a s^{1/12}$ as well.

We will now analyze the scattering occuring on the IR brane, the details
of which are in the Appendix. The wave profile is 
\be
\Phi(r,y)= R_s l 
e^{-2|y|/l} \int_0^{\infty} dq J_0(qr) \frac{I_2(e^{-|y|/l}lq)}
{I_1(lq)}
\ee
and we find that we can use the contour integration over the complex q plane 
to calculate 
\be
f= \Phi(r;y=0)
\simeq R_s \sqrt{\frac{2\pi l}{r}}\sum_n \frac{j_{1,n}^{-1/2}J_2(j_{1,n})
}{a_{1,n}}e^{-\bar{q}_n r}
\simeq R_s \sqrt{\frac{2\pi l}{r}} C_1 e^{-\bar{q}_1r}
\ee
where $\bar{q}_n=  j_{1,n}/l$ are poles of the integral in the complex 
momentum plane given by the zeroes of the Bessel function $J_1$, $j_{1,n}$.
In (\ref{trapped}) from the general formalism, i.e.
\be
\Psi= f+a y +g\frac{y^2}{2}+...=C
\ee
we have (notice the sign difference in g, due to the IR brane metric
(\ref{IRmetric}) vs. the UV brane metric (\ref{UVmetric}) used in section 3)
\be
a=\zeta_1, \;\;\; g= \partial_y^2\Phi
|_{y=0}-\frac{4}{l} \zeta_1,= -f\bar{q}_1^2
-\frac{4a}{l}
\ee
Again, as in the AdS case, we find two solutions, one that would be there 
if we had a 4d scattering, and a larger trapped surface that aapears only when 
we have 5d scattering. In the Appendix we treat in detail the case of the 
4d scattering, for completeness.
Taking instead the larger trapped surface,
\be
a=-3\frac{f \; r^2}{l^3}
\ee
we get the continuity condition at y=0
\be
(\frac{|f|\bar{q}_1}{2})\frac{|\alpha |r}{\bar{q}_1l^2}=1
\Rightarrow \frac{R_s\bar{q}_1}{2}
\sqrt{\frac{2\pi l}{r}}\frac{|\alpha |C_1}{\bar{q}_1 l}\frac{r}{l}
 e^{-\bar{q}_1r}=1 
\ee
It has a solution that is approximately
\bea
&&r_H\simeq \frac{1}{2\bar{q}_1}ln (\frac{\bar{A}}{2\bar{q}_1})\nonumber\\
&& \bar{A}= (\frac{R_s \bar{q}_1}{2})^2 \frac{2\pi }{l} (\frac{3C_1}{
\bar{q}_1 l})^2 
\eea

We can see that there is a solution by considering the function
\be
\tilde{g}(r)= r-\bar{A}r^2 e^{-2\bar{q}_1r}\Rightarrow 
\tilde{g}'(r)= 1+(\bar{q}_1-\frac{1}{r})2\bar{A}r^2 e^{-2\bar{q}_1r}
\ee
and the solution for $r_H$ is given by $\tilde{g}(r)=0$. Since 
$\tilde{g}'(r)>0$ if $r>1/\bar{q}_1$ and 
\be
\tilde{g}(\frac{1}{\bar{q}_1})= \frac{1}{\bar{q}_1}[1-\frac{R_s^2}{l^2}
\frac{18 \pi C_1^2}{4 e^2 j_{1,1}}]<0
\ee
for sufficiently large $R_s$, there will be a solution. 

At nonzero impact parameter, we get the equation
\be
(\frac{|f|\bar{q}_1}{2})^2(\frac{|\alpha |r}{\bar{q}_1l^2})^2(1-\frac{b^2}{
2r^2})=1
\ee
Its solution is the zero of the function
\be
g(r)= r-(1-\frac{b^2}{2r^2})\bar{A} r^2 e^{-2\bar{q}_1r}
\ee
And now we have an analysis that is a bit more involved. 
\be
g'(r)= 1+2\bar{q}_1(r^2-\frac{b^2}{2})\bar{A}e^{-2\bar{q}_1r}-2r\bar{A}
e^{-2\bar{q}_1r}
\ee
We are however only interested in the maximum value $b_{max}$ for which 
$g(r)=0$ has a solution. 

If $r^2<b^2/2$, then $g(r)>0$, and $g(0)= b^2\bar{A}/2,\; g(b/\sqrt{2})
=b/\sqrt{2}$. To have a solution of $g(r)=0$, we need a minimum, $g'(r_1)
=0$, with $g(r_1)<0$, so necessarily $ r_1^2>b^2/2$. 

But if $b^2/2\ll r^2$, the second term in $g'(r)$ is larger than the third
(as $\bar{q}_1\sim 1/l \gg r$), so $g'(r)>0$. Thus we need instead
\be
\frac{b^2}{2r_1^2}=1-\epsilon\simeq 1
\ee
so if $g(r_1)<0$ (but close to zero, so that the 1 is negligible in $g'(r)$
and we can use $g'(r_1)=0$), we get 
\be
g(r_1)\simeq r_1(1-\frac{\bar{A}}{\bar{q}_1}e^{-2\bar{q}_1r})
\ee
At $b=b_{max}, \; g(r_1)=0$, so 
\bea
&&
r_1\simeq \frac{1}{2\bar{q}_1}ln (\frac{\bar{A}}{\bar{q}_1}), b^2\simeq 2r_1^2
\nonumber\\
&&\Rightarrow b_{max}(s)\simeq \sqrt{2}\frac{1}{2\bar{q}_1}
 ln (\frac{\bar{A}}{\bar{q}_1})=\frac{\sqrt{2}}{\bar{q}_1}ln [ R_s \bar{q}_1
(\frac{3\sqrt{\pi}}{\sqrt{2}j_{1,1}^{3/2}})]
\eea
where $R_s=\sqrt{s}G_4, G_4=1/(lM_{P,5}^3)$. 
Thus we get 
the same formula for $r_H(m)$ as \cite{gid} (modulo different constants), 
and thus the classical cross section in the gravity dual
for scattering on the IR brane is of the expected $ln^2 s$ functional form
\be
\sigma =\pi b_{max}^2(s)
\simeq 2 \pi [ \frac{1}{\bar{q}_1}ln (K_1 \sqrt{s}\bar{q}_1G_4)]^2
\ee

\section{ QCD scattering regimes; the Froissart bound}

Finally, now that we have done all the calculations needed for the scattering
in the AdS dual, let's put everything together. In the gravity dual, 
the scattering happens inside AdS or, in the Froissart regime, effectively 
on the IR brane.

In \cite{kn}, we have analyzed 't Hooft scattering in AdS, and we have 
found that we can calculate analytically 
the scattering amplitude in two limits: 
\bea
&& {\cal A}_{AdS}\simeq \frac{G_4l}{2\pi}\frac{s}{\sqrt{t}}
e^{\frac{3y-y_0}{2l}}e^{-\sqrt{t}l(e^{y/l}-e^{y_0/l})}, y\neq y_0<l, 
r\ll l , G_4 s \ll 1\nonumber\\
&& {\cal A}_{AdS}\propto G_4 l^6 s t^2 ln (t),\;\;\; r\gg l
\eea
to be compared with the result in flat D dimensions (which can 
be used in the case $y=y_0, r\ll l$, for instance)
\be
{\cal A}_{AdS}\sim \frac{G_4s}{t}
\ee

All cases still give ${\cal A}\sim G_4s$, only the t behaviour is modified,
so
\be
\frac{d\sigma_{AdS}}{d^2k}= \frac{4}{s}\frac{d\sigma_{AdS}}
{d\Omega}\sim (G_4s)^2
\ee

Since we have a well defined amplitude, we can use the Polchinski-Strassler
formula (\ref{ps}) to relate to a QCD amplitude. Of course we have to remember
that the formula is valid only for the large r region, away from the IR 
cut-off, where we can approximate the metric with AdS and the wavefunction 
dependence of r with the power law $r^{-\Delta}$. But unfortunately, 
in all the cases of interest, studied in (\ref{behav}), the main contribution 
to the integral  comes from the region of large $\nu$ ($\nu\simeq \nu_{max}=
\hat{\alpha}'t$), corresponding to the region of small r, $r\simeq r_{min}$.

This is however not so bad as it seems, since this IR modification 
will translate only in the modification of the factorized t behaviour (coming 
from the integration over $\nu$), while the dependence of s comes from the 
s dependence of the AdS amplitude. We can check this by looking at the 
formula (\ref{ps}). As we will be interested only in the leading s behaviour
and not the t dependence (be it multiplicative or not), our results will 
still be valid (see also the discussion further of the original 
Polchinski-Strassler paper \cite{ps} later on, especially around 
(\ref{regge})).

We also see that in all the cases of interest treated in (\ref{behav}), 
the effect of the AdS integration is only to change the overall normalization,
as well as the subleading behaviour. The leading behaviour is kept the same.

In conclusion, for the case at hand, of the 't Hooft scattering in AdS, 
where ${\cal A}_{AdS}\sim G_4 s$, by using the PS formula (\ref{ps}) we 
will still have the same s dependence 
\be
{\cal A}_{QCD}\sim \hat{G}_4 s \Rightarrow
\frac{d\sigma_{QCD}}{d^2k}= \frac{4}{s}\frac{d\sigma_{QCD}}
{d\Omega}\sim |{\cal A}_{QCD}|^2 \sim (\hat{G}_4s)^2
\ee

As we mentioned, the relation between AdS and QCD energy scales is: 
 $\sqrt{\alpha '\tilde{s}}|_{AdS}\leq \sqrt{\hat{\alpha}' s}
|_{QCD}$ and $\sqrt{\alpha '\tilde{t}}|_{AdS}\leq \sqrt{\hat{\alpha}' t}
|_{QCD}$. Here 
\be
\hat{\alpha}'=\frac{1}{\Lambda^2\sqrt{g_sN}}
\ee
where $\Lambda$ is the mass gap (the lightest glueball state). And the minimum
r in the cut-off AdS is $r_{min}= R^2\Lambda$, $R\equiv l$ in our notation. 

We can rescale the 4d coordinates x such as $r_{min}= R\Rightarrow \hat{\alpha}
'=\alpha '$. The Planck scales are 
\bea
&&M_P(AdS)= \frac{g_s^{-1/4}}{\sqrt{\alpha'}}\leftrightarrow \hat{M}_P
=\frac{g_s^{-1/4}}{\sqrt{\hat{\alpha}'}}=N^{1/4}\Lambda
\nonumber\\&&
\Rightarrow \frac{\sqrt{\tilde{s}}}{M_P}\leq \frac{\sqrt{s}}{\hat{M}_P}
\;\;\; \hat{G}_4\equiv \hat{M}_P^{-2}= \frac{1}{\sqrt{N}\Lambda^2}
\eea

So from 't Hooft scattering in AdS we get a Rutherford-type behaviour in 
QCD as well, except with the effective coupling $\hat{G}_4s$. 

This behaviour is universal, and if we interpret it as single particle 
exchange between the gauge invariant Yang-Mills states (glueballs), we 
see that the particle being exchanged must also be colourless. 

So in the energy regime $\hat{G}_4 s<1$, but at energies larger than the 
mass gap, i.e. for 
\be
\Lambda < \sqrt{s}< \frac{1}{\sqrt{\hat{G}_4}}=\Lambda N^{1/4}
\ee
colourless states should obey Rutherford scattering behaviour, with 
a universal colourless single-particle exchange, with universal 
effective coupling $\hat{G}_4 s$. In the case of real QCD (N=3), the energy 
regime is nonexistent, thus we can't draw any lessons from experiments, and 
the interest in this regime is therefore just theoretical. 

However, as the universal coupling was obtained from a spin 2 exchange in AdS
(graviton), it is natural to assume that the same thing happens in the gauge 
theory, namely a universal spin 2 colourless particle is exchanged in this 
regime that behaves as a graviton. There is a natural candidate for such 
a composite ``particle'', namely the gauge theory dual to the graviton, 
the energy-momentum tensor. It is not obvious why in this energy regime 
the energy-momentum tensor should have a graviton-like universal 
coupling to all colourless states, nor why graviton exchange in the bulk 
should be dual to energy-momentum tensor exchange on the boundary, but 
this is the only plausible candidate.

This is an elastic scattering in AdS, and it is an elastic 
cross section in QCD. It could then maybe be possible to detect this 
elastic $\sigma$ even at 
higher energies, when the amplitude is mostly inelastic.

As we go even higher in energies ($\alpha 's \gg 1$), 
we will start observing the Regge behaviour
noted by Polchinski and Strassler \cite{ps}. Then in AdS we need to 
use the string Virasoro-Shapiro amplitude for massless external states
(s+t+u=$\sum m_i^2=0$) 
\bea
&&{\cal A}= [\prod _{x=s,t,u} \frac{\Gamma(-\alpha ' \tilde{x}/4)}{
\Gamma(1+\alpha ' \tilde{x}/4)}]K(\tilde{p}\sqrt{\alpha '})
=\frac{\Gamma(\alpha 's/4+\alpha 't/4)}{\Gamma(1+\alpha 's/4)}\times
\nonumber\\&&
\times \frac{\Gamma(-\alpha ' s/4)}{\Gamma(1-\alpha 's/4-\alpha 't/4)}
\frac{\Gamma(-\alpha ' t/4)}{\Gamma(1+\alpha 't/4)} K(\tilde{p}\sqrt{\alpha '})
\eea
becoming in the small angle  ($s\gg t$) region of interest
\be
{\cal A}=(\alpha ' s/4)^{\alpha 't/4-1}(-\alpha ' s/4)^{\alpha 't/4-1}
\frac{\Gamma(-\alpha ' t/4)}{\Gamma(1+\alpha 't/4)} s^4 ({\rm pol.\;\; tensors}
)\sim (\alpha ' s)^{\alpha 't/2+2}
\frac{\Gamma(-\alpha ' t/4)}{\Gamma(1+\alpha 't/4)}
\ee

Regge behaviour in QCD was found to correspond to Regge behaviour in AdS
(to the flat space Regge behaviour of this VS amplitude, to be exact). The 
approximation used though is the same approximation that we needed to use 
for the 't Hooft scattering. Namely, in general the $\nu$ integral is 
dominated by (using stationary phase approximation)
\be
\nu_0=- \frac{\Delta -4}{ln (s/|t|)}
\ee
But Regge behaviour is obtained in the case that $\nu_0> 
\nu_{max}=\hat{\alpha}'t$, when
the integration is  dominated by the upper limit of the integral, 
corresponding to r close to $r_{min}$. 

But as in the case of 't Hooft scattering, we only need to assume that the 
maximum of the integral is still outside the region of integration, so that 
we can approximate the integral with its upper limit. Presumably the 
existence of the mass gap is enough to satisfy this requirement. 
Then the factorized 
t dependence will be modified, but the s behaviour will still be of 
Regge type:
\be
{\cal A}\sim (\hat{\alpha} 's)^{2+\hat{\alpha}'t/2}
\label{regge}
\ee
Finally, for the elastic  amplitude, \cite{ps} find yet another 
regime. If $\nu_{max}> \nu_0$ (to be rigorous we would need $\nu_{max}\gg
\nu_0$, corresponding to $ r_{scatt}\gg r_{min}$), that is if
\be
\hat{\alpha }' t> \frac{(\Delta -4)}{ln (s/|t|)}
\ee
one obtains 
\be
{\cal A}\sim s^2 |t|^{-\Delta/2}[ln (s/|t|)]^{1-\Delta/2}
\ee

Let us now turn to energies higher than the Planck scale in AdS, 
$\tilde{s}>M_P^2\Rightarrow s> \hat{M}_P^2$, when we will produce 
black holes in the scattering. There are 3 dimensionless
parameters characterizing the scattering in AdS, $R_s/l, G_4s$ and $l M_{P,5}$,
so let's first derive their relations to QCD variables. 

Since $M_{P,5}= N^{1/4}/l$ in our notation, then 
\be
R_s= G_4 \sqrt{\tilde{s}}= \frac{\sqrt{\tilde{s}}}{lM_{P,5}^3}
\leq (\frac{1}{lM_{P,5}^2})\frac{\sqrt{s}}{\hat{M}_P}|_{QCD}
\ee
so that 
\be
\frac{R_s}{l}\leq \frac{1}{\sqrt{N}}(\frac{\sqrt{s}}{\hat{M}_P})|_{QCD}
\ee

Also, since $G_4\tilde{s}= R_s^2/G_4$, we get
\be
G_4\tilde{s}\leq N^{-1/4}(\frac{\sqrt{s}}{\hat{M}_P})^2|_{QCD}
\ee

Equivalently, in terms of $\hat{\alpha}'$, the two parameters satisfy
\be
\frac{R_s}{l}\leq \frac{g_s^{1/4}}{\sqrt{N}}\sqrt{\hat{\alpha}'s}|_{QCD},
\;\; G_4 \tilde{s}\leq \frac{g_s^{1/2}}{N^{1/4}}(\hat{\alpha}'s)|_{QCD}
\ee

Finally,  $lM_{P,5}=  N^{1/4}\gg 1$, so one first 
reaches the AdS scale $1/l$, then $M_P\equiv M_{P,5}$, and then the 
scale $E_R= M_P(lM_P)^{d-3}$, when the black hole size is comparable 
with the AdS size. 

Note that the dimensionality of the gravity dual plays an important role. 
Most of the calculations in this paper were done assuming that there is 
only $AdS_5$ and forgetting about the compact space. In the case of 
't Hooft scattering, we argued that the relevant behaviour was independent 
of the dimensionality of the space. In the case of black hole creation, 
the dependence is more important.

Before the size of AdS becomes important, black hole formation can be 
approximated as being in flat space. In that case, we have calculated in 
\cite{kn} a lower limit on the maximum impact parameter that creates a 
black hole in d dimensions,
\bea
&&b^2_{max}\leq 2[\frac{\alpha}{D-2}]^{\frac{1}{D-3}}\frac{D-3}{D-2}= 
\frac{2(\epsilon r_H)^2}{ [D-2]^{\frac{D-2}{D-3}}}(D-3)\nonumber\\
&&r_H=[\frac{16\pi G \sqrt{s}}{(D-2)\Omega_{D-2}}]^{\frac{1}{D-3}}
\nonumber\\
&&\epsilon=[\frac{(D-2)\Omega_{D-2}}{4\Omega_{D-3}}
]^{\frac{1}{D-3}}
\eea
In any case, we see that $b_{\max}(s)\simeq a s^{\frac{1}{2(D-3)}}= a 
s^{\beta}$ (a= constant). In section 2.1 we have used a simple black disk 
model to create an imaginary elastic scattering amplitude that was substituted
in the Polchinski-Strassler formula (\ref{ps}). We have derived the 
forward (t=0) imaginary part of the amplitude, giving us the total 
QCD scattering cross-section in this regime
\be
\sigma_{QCD}= \frac{\pi a^2 K}{2\beta +\Delta/2 -1}(\frac{\hat{\alpha}'s}
{\alpha '})^{2\beta}
=\frac{\pi a^2 K}{1/(D-3) +\Delta/2 -1}(\frac{\hat{\alpha}'s}{\alpha '})
^{\frac{1}{D-3}}
\label{bhprod}
\ee
As we have explained in section 2.1, this is of the type 
\be
(K a^{\frac{2}{1+2\beta}}) \times fct. (\frac{a^{\frac{2}{1+2\beta}}}{\alpha '
}, \hat{\alpha '} s, \hat{\alpha '} t, \Delta)
\ee
and here $a^{2/(1+2\beta)}/\alpha '$ is a number depending on the gravity 
dual (dimension D, Newton constant G), whereas $Ka^{2/(1+2\beta)}$ is a 
dimension -2 constant that also depends strongly on the details of the IR 
cut-off (through K). 

We see that the higher the dimensionality, the smaller the dependence on s.
For d=5 we have $\sigma \sim s^{1/2}$, whereas for d=10 we have $\sigma
\sim s^{1/7}$.

At even higher energies ($\sqrt{s}>E_R$), we will start feeling the 
effects of the AdS size. We have calculated in section 4 that the 
maximum impact parameter for black hole formation is at least equal to
\be
b_{max}=7^{-1/12}\sqrt{\frac{12}{7}}l e^{y_0/l}[\frac{R_s}{le^{y_0/l}}]^{1/6}
=\frac{1}{7^{1/12}2^{2/3}}\sqrt{\frac{12}{7}}r_{max}
\ee
where $R_s=2G_4 \sqrt{s}$, so that $b_{max}\simeq a' s^{1/12}$.

Then using the same black disk model to substitute in the P-S formula, we 
get 
\be
\sigma_{QCD}= \frac{\pi {a'}^2 K}{2\beta +\Delta/2 -1}(\frac{\hat{\alpha}'s}{
\alpha '})^{2\beta}
=\frac{\pi {a'}^2K}{1/6 +\Delta/2 -1}(\frac{\hat{\alpha}'s}{\alpha '})
^{\frac{1}{6}}
\label{adsscat}
\ee

But as in the previous cases, the P-S formula shows that 
the main integration region is near the cut-off $r_{min}$, which corresponds 
to the IR brane in the 2-brane RS model. But in that case we can't approximate
the scattering as being in AdS, since the 4d size of the black hole 
formed is comparable to the AdS scale.
As $\rho\gg l$ the horizon will stretch over a size $\Delta y>l$, and if 
$y\simeq y_{IR}$ it would look as if the black hole is approximately on the 
IR brane. So the approximation of A-S in AdS will break down and instead the 
good approximation would be the two A-S shockwaves being on the IR brane.

So it is not even clear that there is an intermediate regime of the type 
in (\ref{adsscat}), but it is clear that the cross section will begin 
flattening out, finally to settle into the final behaviour, corresponding
to scattering on the IR brane.

In the second part in section 4, we have calculated that for A-S scattering
on the IR brane, we get a
maximum impact parameter for black hole formation that is at least equal to
$b_{max}=\sqrt{2}r_1$, with 
\be
r_1=\frac{1}{2\bar{q}_1}ln (\frac{\bar{A}}{\bar{q}_1});\;\;\;
\bar{A}= (\frac{R_s \bar{q}_1}{2})^2 \frac{2\pi }{l} (\frac{3C_1}{
\bar{q}_1 l})^2 
\ee

In that calculation  we have used the metric
\be
ds^2= e^{2|y|/l}d\vec{x}^2 +dy^2= \frac{r^2}{l^2}(\frac{l^2}{r_{min}^2}
d\vec{x}^2)+\frac{l^2}{r^2}dr^2
\ee
so to go back to the real coordinates we substitute 
\be
\tilde{\rho}(real)=\frac{l}{r_{min}}\rho(used)=\frac{\rho (used)}{l\Lambda}
\ee

Thus
\be
b_{max}(s) \simeq \frac{1}{l\Lambda}
\sqrt{2}\frac{1}{2\bar{q}_1}
 ln (\frac{\bar{A}}{\bar{q}_1})=\frac{\sqrt{2}}{\bar{q}_1}ln [ R_s \bar{q}_1
(\frac{3\sqrt{\pi}}{\sqrt{2}j_{1,1}^{3/2}})]
\ee
and $R_s=\sqrt{s}G_4$, $G_4= 1/(l M_{P,5}^3)$. As advocated, we get 
the same formula for $r_H(m)$ as in \cite{gid}, modulo different constants.

The gravitational cross section is 
\be
\sigma \simeq \pi [\frac{\sqrt{2}}{l\Lambda}
 \frac{1}{\bar{q}_1}ln (K \sqrt{s}\bar{q}_1G_4)]^2
\label{frois}
\ee
and $\sigma_{QCD}\sim \sigma$, as we argued.

Thus as expected, the final behaviour of the QCD scattering amplitude 
corresponds to the Froissart unitarity bound. We have obtained the same 
behaviour that Giddings \cite{gid} has proposed, but in a more rigorous 
setting. Let us compare to the calculation in \cite{gid}. There, the 
``Newton potential'' $h_{00}$  was calculated in linearized gravity ,obtaining
\be
h_{00}\sim e^{\frac{dy}{2l}}\int \frac{d^{d-1}\vec{p}}{(2\pi)^{d-1}}
e^{i\vec{p}\vec{x}}\frac{J_{d/2}(ip l e^{y/l})}{pJ_{d/2-1}(ipl)}
\ee
which was then used for an estimate of the horizon size by $h_{00}\sim 1
\Rightarrow r=r_H$, and a geometric cross-section approximation 
 $\sigma \simeq \pi r_H^2$ was used for the black hole.

In our case, we obtain the exact A-S shockwave solution on the IR brane, 
which is 
\bea
\Phi (r,y)
&=& \frac{R_sl}{2\pi}
e^{-\frac{d|y|}{2l}}\int \frac{d^{d-2}\vec{q}}{(2\pi)^{d-2}}e^{i\vec{q}
\vec{x}}\frac{I_{d/2}(e^{-|y|/l}lq)}{qI_{d/2-1}(lq)}\nonumber\\
&=&\frac{R_s l}{(2\pi)^{\frac{d-4}{2}}}\frac{e^{-\frac{d|y|}{2l}}}{r^{\frac{d-4
}{2}}}\int_0^{\infty} dq q^{\frac{d-4}{2}}J_{\frac{d-4}{2}}(qr)\frac{I_{d/2}
(e^{-|y|/l}lq)}{I_{d/2-1}(lq)}
\eea
and since $J_{\nu}(ix)=I_{\nu}(x)$ and y is negative, is very similar to the
Giddings case, except that for us the integration is only over d-2 transverse 
coordinates (as for a massless particle), whereas for \cite{gid} 
the integration 
is over d-1 transverse coordinates, as appropriate for a massive particle.
The $h_{00}$ in \cite{gid} was obtained also 
from a wave equation with sources on the IR brane, except there the
 source was a static mass (black hole), whereas for us it is a photon. 

At y=0, one obtains similar behaviours,
\be
h_{00}\simeq \frac{1}{2\pi r} \sum_n e^{-\bar{q}_n r} \frac{J_2(j_{1,n})}{
l a_{1,n}}\simeq \frac{1}{2\pi r} e^{-\bar{q}_1r}\frac{J_2(j_{1,1})}{l
a_{1,1}}
\ee
versus
\be
\Phi(r,y=0)\simeq R_s \sqrt{\frac{2\pi l}{r}}\sum_n \frac{j_{1,n}^{-1/2}
J_2(j_{1,n})}{a_{1,n}}e^{-\bar{q}_n r}
\simeq R_s \sqrt{\frac{2\pi l}{r}} C_1 e^{-\bar{q}_1r}
\ee
and in both cases the fact that allows the logarithmic behaviour of $r_H$
is the exponential $e^{-\bar{q}_1r}$, itself coming from the presence of the
pole in the momentum space integrand. In our case, we have the advantage of 
the scattering picture, in which we can calculate directly 
the cross section for black hole creation.

Finally, in \cite{kn} we have addressed the issue of string corrections 
to the scattering. This is very relevant for the case of QCD, since 
string $\alpha '$ and $g_s$ corrections in AdS  translate into $1/N$ and
$1/(g^2_{YM}N)$ corrections in the gauge theory, bringing us closer to the 
case of QCD. So it is important to realize their effect. 

In \cite{kn}, string corrections were analyzed using a formalism of 
Amati and Klimcik \cite{ak} (as well as using the formalism in \cite{acv93}, 
based on the action in \cite{lipatov}, but we will not describe it), 
in which one obtains a string-corrected 
A-S metric. These corrections depend the dimensionless ratio (in d=4; 
here $Y=\alpha ' log(\alpha 's)$)
\be
\frac{R_s^2}{Y}=\frac{g^2}{log (\alpha ' s)}\frac{g^2\alpha' s}{(4\pi)^2}
\ee

When this parameter is large, the shockwave is approximately A-S, 
with exponentially small corrections, namely
\be
\Phi(b)= -\frac{g^2\sqrt{s}}{4\pi}\alpha '(2\;log \; \frac{b}{R_s}- 
e^{-\frac{b^2}{4Y}}
(\frac{b^2}{4Y})^{-1}+...)= -R_s(2\;log \; \frac{b}{R_s}- e^{-\frac{b^2}{4Y}}
(\frac{b^2}{4Y})^{-1}+...)
\ee
and the maximum impact parameter for black hole creation also 
increases, but with exponentially small corrections: 
\be
B_{max}=\frac{R_s}{\sqrt{2}}(1+e^{-R_s^2/(8Y)})
\ee

When $R_s^2/Y\ll 1$, the shockwave is not of A-S type
\be
\Phi(b)=-2R_s(\frac{1}{D-4}-\frac{b^2}{4Y(D-2)}+...)
\ee
and we can show that $B_{max}\sim \sqrt{Y}\gg R_s$, (but 
we can't compute the actual value), so one has a huge increase in the 
cross section. 

So string corrections can be quite important, but in the case of $s\rightarrow
\infty$ (the Froissart unitarity bound), $R_s^2/Y\gg 1$, and the corrections 
discussed here are exponentially small. Thus the Froissart unitarity bound 
is unaffected, as expected.

While this calculation was in flat d=4, not in the gravity dual, and string 
corrections were described using a very simple model, the smallness 
of the corrections, together with the fact that we do obtain what we expect, 
namely the Froissart bound, makes us believe that we are correctly describing 
a QCD phenomenon.

\section{Conclusions}

In this paper we have analyzed high energy QCD scattering in the small angle 
region $s\gg t$. We have applied the high energy gravitational 
scattering calculations in \cite{kn} to the simple model of QCD gravity 
dual used in \cite{ps}. Namely, for high energy QCD many of the 
observed features can be deduced from just a AdS dual cut-off in the 
IR (and maybe in the UV), giving a 2-brane RS model. 
The gravitational scattering was analyzed using a shockwave analysis. 
At energies smaller than the Planck scale, the scattering of two massless 
particles is described by null geodesics propagating in the shockwave 
background ('t Hooft scattering). At energies higher than the Planck scale,
we need to take two A-S shockwaves, and black holes are being formed in the 
future of the collision, for an impact parameter less than a $b_{max}(s)$.

't Hooft scattering corresponds in the gauge theory to a very restrictive 
regime ($\Lambda <\sqrt{s}<\Lambda N^{1/4}$) that is too restrictive for 
real QCD (N=3). For N large though, we obtained a Rutherford-type scattering 
with effective coupling $\hat{G}_4s$, implying a universal
single-particle exchange.
We conjectured that this comes from an exchange of a universal,
graviton-like spin 2 ``particle'' being exchanged, most likely the 
energy momentum-tensor (the dual of the graviton).
As the cross-section is elastic, one could however maybe detect this elastic 
$\sigma$ even at higher energies in the gauge theory, when the amplitude 
is mostly inelastic, thus being of possible relevance to real QCD. 

At higher energies, Regge behaviour sets in, as described by \cite{ps}. 
Regge behaviour of the flat space Virasoro-Shapiro amplitude translates 
directly into Regge behaviour in QCD. 

At even higher energies, black hole production sets in inside the gravity 
dual, and it can be described as if happening in an approximately flat 
d-dimensional space, giving the power law behaviour in (\ref{bhprod}), 
namely $\sigma_{QCD}\sim s^{1/(d-3)}$. 

There is a possible transition region 
when the size of AdS becomes important, and the scattering in the gravity 
dual created black holes inside AdS, giving again a power law behaviour
(\ref{adsscat}), namely $\sigma_{QCD}\sim s^{1/6}$.

Finally, in the last energy regime
most of the scattering happens on the IR brane, and the size of 
the surrounding AdS is important. We obtain a shockwave solution that 
has the same exponential behaviour $h\sim e^{-\bar{q}_1r}$ as the 
linearized black hole solution used by Giddings \cite{gid}. The scattering
of two such shockwaves gives a scattering cross section saturating the 
Froissart bound (\ref{frois}).

We have looked at string corrections to the scattering of two modified 
shockwaves, and we have found that in the simple Amati-Klimcik model 
used in \cite{kn}, and in flat d=4, we obtain exponentially small 
corrections in the Froissart limit $s\rightarrow \infty$. This makes us 
believe that even in the case of real QCD, when N and $g^2_{YM}N$ are 
finite, thus we have large string corrections in the gravity dual, 
our calculation of the Froissart bound still applies.

{\bf Acknowledgements} We would like to thank Matt Strassler for a lot of 
useful discussions on his work, and also thank Antal Jevicki and Carlos Nunez.
This research was  supported in part by DOE
grant DE-FE0291ER40688-Task A.

\newpage

{\Large\bf{Appendix A. Trapped surface calculations}}

\renewcommand{\theequation}{A.\arabic{equation}}
\setcounter{equation}{0}

Let's apply the general formalism to calculate the shape of the trapped 
surface in AdS.

The continuity equation at b=0 would be, for $\Psi=\Phi$, 
\be
(\partial_i \Phi)^2+e^{-2y/l}(\partial_y \Phi)^2=4
\ee
which for $e^{y/l}l/r\gg 1$ gives
\be
e^{\frac{3y-y_0}{l}}(\frac{\bar{C}}{4l}
)^2[\frac{1}{[r^2+l^2(e^{y/l}-e^{y_0/l})^2]^2}
+\frac{9}{4l^2 e^{2y/l}[r^2+l^2(e^{y/l}-e^{y_0/l})^2]}]=1
\ee
(where $\bar{C}=2R_s l^2$), since
\be
\Phi\simeq
 \frac{\bar{C}e^{\frac{3y-y_0}{2l}}}{2l}\frac{1}{\sqrt{r^2+l^2(e^{y/l}
-e^{y_0/l} )^2}}
\ee
At $y=y_0$ we get
\be
(e^{y_0/l}\frac{R_sl}{2r^2})^2(1+\frac{9}{4}\frac{r^2}{l^2e^{2y_0/l}})=1
\ee
where as we can see the second term (coming from $(\partial_y\Phi)^2$) is 
a small correction.

For $l\ll r$ we have 
\be
\Phi= \frac{\bar{C}l^4}{r^6} e^{\frac{2}{l}(2y+y_0)}
\ee

That implies 
\be
[\frac{6R_s l^6}{r^7}e^{\frac{2}{l}(2y+y_0)}]^2[1+\frac{4}{9} \frac{r^2}{l^2
e^{2y/l}}]=1
\ee
However, now the second term (coming again from $(\partial_y\Phi)^2$)
is dominant. 

But we still need to modify $\Phi$: $\Psi=\Phi +\zeta$, as in the RS case. 
Using the formulas from the RS case described in section 3, but remembering 
that now we expand in $\epsilon= y-y_0$, not in y, we get 
\be
\Psi= C= f +a \epsilon +\frac{\epsilon^2}{2} g+...
\ee
where 
\be
f= \Phi|_{y=y_0},\;\;\; a= \zeta_1+\partial_y \Phi|_{y=y_0}
\ee
Using  
\be
\Phi= K \frac{e^{\frac{d\epsilon}{2l}}}{r^{\frac{d-4}{2}}}\int _0^{\infty}
dq q^{\frac{d-2}{2}}J_{\frac{d-4}{2}}(qr) K_{d/2}(e^{\frac{y_0+\epsilon}{l}} 
lq)I_{d/2}(e^{y_0/l}lq)
\ee
and $ zK_{\nu}'(z)+\nu K_{\nu}(z)=-z K_{\nu -1}(z)$, we get in d=4
\be
\partial_{\epsilon}\Phi =-\frac{Ke^{\frac{2\epsilon}{l}}}{l}\int_0^{\infty}
dq q J_{0}(qr) (e^{\epsilon/l}e^{y_0/l}lq)K_1(e^{\epsilon/l}e^{y_0/l}lq)
I_2(e^{y_0/l}lq)
\ee
Then using $I_2(z)=-2I_1(z)/z+ I_0(z)$ and 
\bea
&&\int_0^{\infty}x J_0(x) K_1(ax) I_1(ax)= \frac{1+2a^2-\sqrt{1+4a^2}}{2a^2
\sqrt{1+4a^2}}\nonumber\\&& 
\int_0^{\infty} dx x^2 J_0(x) K_1(ax)I_0(ax)
=2a(1+4a^2)^{-3/2}
\eea
we get 
\be
\partial_{\epsilon}\Phi|_{\epsilon=0}= \frac{Ke^{-2y_0/l}}{l^3}[
\frac{6e^{4y_0/l}l^4/r^4+6 e^{2y_0/l}l^2/r^2+1}{(1+4e^{2y_0/l}l^2/r^2)^{3/2}}
-1]
\ee
At $e^{y_0/l}l/r\ll 1$, using $K= \bar{C}e^{2y_0/l}=2R_s l^2 
e^{2y_0/l}$, we have
\be
\partial_{\epsilon}\Phi|_{\epsilon=0}\simeq 4\bar{C}e^{6y_0/l}\frac{l^3}{r^6}
=4 \frac{\Phi|_{\epsilon=0}}{l}
\ee
Then also 
\be
\partial^2_{\epsilon}\Phi|_{\epsilon=0}=\frac{8}{l^2}\Phi|_{\epsilon=0}
+Ke^{2y_0/l} \int_0^{\infty} dq q^3 J_0(qr) K_0(e^{y_0/l}lq)I_2(e^{y_0/l}lq)
\ee
Now however we can only do the remaining integral if the argument is small. 
Namely, one could try using again $I_2(z)=-2I_1(z)/z+ I_0(z)$ and 
\be
\int_0^{\infty}x^3 J_0(x) K_0(ax) I_0(ax)=-4\frac{1+2a^2+4a^4}{(1+4a^2)^{3/2}}
\ee
but the $I_1$ integral cannot be done. 
Instead, expand $K_0(ax)I_2(ax)$ and find 
\be
\int_0^{\infty}x^3 J_0(x) K_0(ax) I_2(ax)\simeq 8a^2-96 a^4+o(a^6)
\ee
and thus for  $e^{y_0/l}l/r\ll 1$ we have 
\be
\partial^2_{\epsilon}\Phi|_{\epsilon=0}\simeq 16\cdot 2 R_s \frac{l^4}
{r^6}e^{6y_0/l}
\ee
Thus at $e^{y_0/l}l/r\ll 1$ and $y=y_0$
\bea
&& f=\Phi|_{\epsilon=0}= \bar{C} \frac{l^4}{r^6}e^{6y_0/l}\nonumber\\
&& a=\frac{4}{l}f+\zeta_1\nonumber\\
&& g=\partial^2_{\epsilon}\Phi|_{\epsilon=0}+\frac{4}{l}\zeta_1
=\frac{16 f}{l^2}+\frac{4\zeta_1}{l}
\eea
If a is nonzero, we have to match 
\be
4=f'^2 +a^2e^{-2y_0/l}+\epsilon(2a'f'-2\frac{a^2}{l}e^{-2y_0/l}+2ag 
e^{-2y_0/l})+...
\ee
with 
\be
C=f+a\epsilon+...
\ee
If we put $\zeta_1=0$ we then get 
\be
1=(R_s\frac{l^6}{r^6}e^{6y_0/l})^2\frac{16}{l^2}e^{-2y_0/l}(1+\frac{9}{4}
\frac{l^2}{r^2}e^{2y_0/l}+6\frac{\epsilon}{l}+...)
= (\frac{f}{2})^2 \frac{16}{l^2}e^{-2y_0/l}(1+\frac{9}{4}
\frac{l^2}{r^2}e^{2y_0/l}+6 \frac{\epsilon}{l}+...)
\ee
where now the first term comes from $(\partial_y\Psi)^2$, to be matched 
with 
\be
C^2=f^2(1+\frac{8\epsilon}{l}+...)
\ee
which doesn't work. The next try is to put a nonzero $\zeta_1$ that still 
keeps $a$ nonzero. If it keeps it of the same order as the first term, we will 
have 
\be
a=-\frac{\alpha}{l} f
\ee
which implies we have to match
\be 
1= (\frac{f}{2})^2 \frac{\alpha^2}{l^2}e^{-2y_0/l}(1+6
\frac{\epsilon}{l}+...)
\ee
with 
\be
C^2=f^2(1-\frac{2\alpha\epsilon}{l}+...)
\ee
This happens to have a solution, $\alpha=-3$! But let us see if it is 
unique. We could also have $a=0$ to this order ($\alpha =0$). Then let's 
assume that a is proportional to the next term in the expansion, namely 
\be
a= \beta \frac{f}{r}e^{y_0/l}
\ee
so that $\zeta_1 = 4f/l+a$ and so $g= 4a/l+...$. Then one finds 
\be
1=(\frac{f}{2r})^2 (6^2+\beta^2)(1+\frac{\epsilon}{l} \frac{6\beta^2}
{6^2+\beta^2})
\ee
to be matched with  
\be
C^2=f^2(1+2\epsilon \frac{\beta}{r}e^{y_0/l}+...)
\ee
and as we see the functional dependence is different. For a 
higher power in the expansion
if
\be
a=\beta \frac{f}{r}e^{y_0/l}(\frac{l}{r}e^{y_0/l})^n,\;\; n\ge 1
\ee
the first equation will be
\be
1=(\frac{6f}{r})^2 \{ 1+\frac{\epsilon}{6l}[ \beta^2 (\frac{l}{r}e^{y_0/l})
^{2n} +2\beta(7+n)(\frac{l}{r}e^{y_0/l})^{n+1}]\}
\ee
whereas the second will be
\be
C^2=f^2(1+2\epsilon \frac{\beta}{r}e^{y_0/l}(\frac{l}{r}e^{y_0/l})^n +...)
\ee
We see that for $n>1$ we have a mismatch, we would need $2(n+7)/6=2$, which 
is impossible, whereas for $n=0$ there is a different functional dependence 
(and for $n<-1$ we can treat it separately and convince ourselves that 
the functional dependence is also different). But for n=1 we have another 
solution, the matching condition is $\beta(\beta +4)=0$, so 
\be
a=-4\frac{f \;l }{r^2}
\ee

Finally, we can't have a=0 exactly, since then we need to match
\be
4=f'^2+y^2(f'g'+g^2 e^{-2y_0/l})+...
\ee
to 
\be
C=f+\frac{y^2}{2}g+...
\ee
and then we have g=0. 

So we have two solutions, $a=3f/l$ and $a=-4fl/r^2$. Which should we choose?
A physical argument shows us what happens. In the second solution, $a$ is 
negligible at $y=y_0$, so we have the same continuity condition that we would 
have if the shockwave scattering was 4 dimensional. But now we are in AdS 
and the warping is very large, so we are at higher energy away from $ 
y=y_0$, therefore we expect that the black hole formed is larger than what 
we would have in the four dimensional case. So whereas the second solution 
describes a trapped surface that would be there even if the space would be 
four dimensional, the second trapped surface is larger and is due to the 
very large warping outside the 4 dimensional slice. 

We might be worried that there is some theorem stating there should only 
be a trapped surface, as the trapped surface problem is similar to the 
Green problem with Neumann boundary conditions in electrostatics. But 
this is not quite so, since now we must also determine the boundary C
from the condition that $(\nabla \Psi)^2=4$ matches $\Psi=$ (arbitrary!) 
constant, together with the solution to the Laplace equation $\Delta (\Psi-
\Phi)=0$, so it is not quite the same Green problem. 

Thus we take the larger of trapped surfaces, with $a=-\alpha f/l$, in 
which case the continuity condition becomes 
\be
\frac{|\alpha| f}{2l}e^{-y_0/l}=1
\ee
so that 
\be
r=r_{max}=l e^{y_0/l}[\frac{3R_s}{le^{y_0/l}}]^{1/6}
\ee
and since $R_s=2G_4 \sqrt{s}$, we have that $r_{max}\sim s^{1/12}$.

Now let us do the same for the case of the wave on the RS IR brane.

We have 
\be
\int d\Omega_{d-3} e^{iqr \cos \theta} =(2\pi) ^{\frac{d-2}{2}}
\frac{J_{\frac{d-4}{2}}(qr)}{(qr)^{\frac{d-4}{2}}}
\ee
which we can apply for 
\be
\int d^{d-2}\vec{q}e^{i\vec{q}\vec{x}}f(q)=\int_0^{\infty} q^{d-3}dq
\int d\Omega_{d-3}e^{iqr\cos\theta}f(q)
\ee
However, in the case of \cite{gid}, 
for the calculation of $h_{00}$, the integral
that one has corresponds formally to d=5 in the above, but it is more useful 
to do the integral in a different way, namely to write
\be
\int_0^{\infty}q^2 dq \int d\Omega_2 e^{iqr\cos\theta}f(q)=\frac{-2\pi i}{r}
\int_0^{\infty} q dq (e^{iqr}-e^{-iqr})f(q)
\ee
and if f is even, i.e. $f(q)= f(-q)$, we can rewrite it as 
\be
\frac{-2\pi i}{r}\int_{-\infty}^{+\infty} q dq f(q) e^{iqr}
\ee

For \cite{gid}, on the IR brane, that is at y=0, the function f is 
\be
f(q)= \frac{J_2(iql)}{iq J_1(iql)}=\frac{I_2(ql)}{qI_1(ql)}
\ee
($I_{\nu}(iz)= i^{-\nu}J_{\nu}(-z)$). More precisely,
\bea
&&h_{00}(y=0)= \frac{1}{(2\pi )^3}\int d^3\vec{p}e^{i\vec{p}\vec{x}}
\frac{J_2(iql)}{iq J_1(iql)}\nonumber\\
&&=\frac{1}{(2\pi)^3}(\frac{-2\pi i}{r})\int_{-\infty}^{+\infty}
dq q e^{iqr} \frac{I_2(ql)}{qI_1(ql)}
\eea

But there is a theorem: For a complex function $\bar{f}(z)$ such that 
$\lim_{z\rightarrow \infty}\bar{f}(z)=0 (Im(z)>0)$, and a real $\sigma >0$, 
we have
\be
\int_{-\infty}^{+\infty} \bar{f}(x) e^{i\sigma x}dx = 2\pi i \sum_{Im(a)>0}
Rez[F, a], \;\;\; F(z)=\bar{f}(z)e^{i\sigma z}
\ee

In Giddings's case \cite{gid},  $\bar{f}(q)= I_2(ql)/I_1(ql)$

We get
\be
h_{00}\simeq \frac{1}{2\pi r} \sum_n e^{-\bar{q}_nr}\frac{J_2(\bar{q}_nl)}{
a_{1,n}}
\ee
where  we have defined $q=i\bar{q}$ and the behaviour of the Bessel function 
near a pole is 
\be
J_1(z)\sim a_{1,n}(z-z_n),\;\;\; z\rightarrow z_n
\ee
The zeroes of $J_1$ are called $j_{1,n}$, so $\bar{q}_n=j_{1,n}/l$. Then
\be
h_{00}\simeq \frac{1}{2\pi r} \sum_n e^{-\bar{q}_n r} \frac{J_2(j_{1,n})}{
l a_{1,n}}\simeq \frac{1}{2\pi r} e^{-\bar{q}_1r}\frac{J_2(j_{1,1})}{l
a_{1,1}}
\ee

Let us come back to our case, of computing $\Phi (r,y)$.
We will first try to do the integral exactly, and see that unfortunately 
we get nonsensical results, and then make an approximation that allows us 
to do the integral. 

The integral we have is 
\bea
&&I= \int_0^{\infty} q dq \int_0^{2\pi} d\theta e^{iqr \cos\theta} f(q) 
= \int_0^{\infty} q dq f(q) \int_{-\pi/2}^{+\pi/2}d\theta (e^{iqr\cos\theta
}-e^{-iqr\cos\theta})\nonumber\\
&& = \int_{-\pi/2}^{+\pi/2}d\theta\int_{-\infty}^{+\infty} q dq f(q)
e^{iqr\cos\theta}
\eea
and we see that in the $\theta$ integration regime $1\geq \cos\theta\geq 0$, 
and 
\be
f(q)=\frac{I_2(ql)}{qI_1(ql)}
\ee
as before, so that 
\be
I= \int_{-\pi/2}^{+\pi/2}d\theta 2\pi i \sum_n e^{-\bar{q}_nr\cos\theta}
\frac{J_2(j_{1,n})}{l a_{1,n}}
\ee
and correspondingly 
\bea
\Phi(r,y=0)&=&\frac{2\pi i}{(2\pi)^2}R_s \sum_n\frac{J_2(j_{1,n})}{a_{1,n}}
\int_{-\pi/2}^{+\pi/2}d\theta e^{-\bar{q}_nr\cos\theta}\nonumber\\
&=&\frac{2\pi i}{(2\pi)^2}R_s \sum_n
\frac{J_2(j_{1,n})}{a_{1,n}}\pi (I_0(\bar{q}_nr)
-L_0(\bar{q}_nr))
\eea
We are interested in the limit $\bar{q}_nr\gg 1$, and 
\be
L_0(z)\sim I_0(z)\sim \frac{1}{\sqrt{2\pi z}}e^z
\ee
at large z, so that is not very useful. We can instead expand the integral 
already and get 
\be
\int_{-\pi/2}^{+\pi/2}d\theta e^{-\bar{q}_nr\cos\theta}=
2 \int_{0}^{+\pi/2}d\theta e^{-\bar{q}_nr\cos\theta}\simeq 
2\int_0^{\epsilon} d\bar{\theta}e^{-\bar{q}_n r\bar{\theta}}
\simeq \frac{2}{\bar{q}_nr}
\ee
so it would seem that there is no exponential behaviour! However, the 
approximations used were contradictory, since we have expanded the previous 
integral about the point where $\cos\theta=0$, which is exactly the point 
where the contour integration theorem doesn't work. 

So we must find an approximation regime when we can do the integral. 

This time we do the angular integral and obtain 
\be
I= 2\pi \int_0^{\infty} dq J_0(qr) \frac{I_2(ql)}{I_1(ql)}
\ee
Since we want to have $r\gg l$, we can use the large argument expansion 
of $J_0$,
\be
J_0(z)\sim \sqrt{\frac{2}{\pi z}}\cos (z-\pi/4)=\frac{1}{\sqrt{\pi z}}
(\cos z +\sin z)
\ee
thus
\bea
&&I\simeq \sqrt{\frac{2}{\pi r}}\int_0^{\infty} dq q^{-1/2} 
\frac{e^{iqr-i\pi/4}+e^{-iqr+i\pi/4}}{2}\frac{I_2(ql)}{I_1(ql)}
\nonumber\\&&=
e^{-i\pi/4}\sqrt{\frac{1}{2\pi r}}\int_{-\infty}^{+\infty} dq q^{-1/2}
e^{iqr}\frac{I_2(ql)}{I_1(ql)}\nonumber\\
&&=e^{-i\pi/4}\sqrt{\frac{1}{2\pi r}}
2\pi \sum_{Im(q)>0}Rez[e^{iqr}q^{-1/2}
\frac{I_2(ql)}{I_1(ql)}, q]
\eea
so that 
\be
\Phi(r,y=0)
\simeq R_s l \sqrt{\frac{2\pi l}{r}}\sum_n \frac{j_{1,n}^{-1/2}J_2(j_{1,n})
}{l a_{1,n}}e^{-\bar{q}_n r}
\simeq R_s \sqrt{\frac{2\pi l}{r}} C_1 e^{-\bar{q}_1r}
\ee

Let us now  calculate the trapped surface shape in the RS background
with 
\be
\Phi (r,y)= R_s l 
e^{-2|y|/l} \int_0^{\infty} dq J_0(qr) \frac{I_2(e^{-|y|/l}lq)}
{I_1(lq)}
\ee
The continuity condition is 
\be
(\nabla \Phi)^2 + e^{2|y|/l}(\partial_y \Phi)^2 =0
\ee
(note the different sign in the exponent, due to the 2-brane RS background).
We have 
\be
\partial_y \Phi= -R_s l e^{-3|y|/l}\int_0^{\infty} dq q J_0(qr) 
\frac{I_1(e^{-|y|/l}lq)}{I_1(lq)}
\ee
 (where we have used $zI_{\nu}'(z)= zI_{\nu-1}(z)-\nu I_{\nu}(z)$)
which means that $\partial_y \Phi|_{y=0}=0$! And then 
\be
\partial_y^2 \Phi|_{y=0}= \frac{R_s}{l} \int_0^{\infty} dq (lq)^2 J_0(qr) 
\frac{I_0(lq)}{I_1(lq)}= R_s l \int_0^{\infty} dq q^2 J_0(qr)
\frac{I_2(ql)}{I_1(ql)}
\ee
(where we have used $zI_{\nu}'(z)= zI_{\nu-1}(z)-\nu I_{\nu}(z)$ as well 
as $I_0(x)= 2I_1(x)/x +I_2(x)$)

Again, as before, the integral is zero in perturbation theory, so we must 
use the contour integral as before, and obtain
\be
\partial_y^2 \Phi|_{y=0}\simeq \frac{R_s l}{\sqrt{2\pi r}}\frac{2\pi i}{
\sqrt{i}}\sum_{Im(q)>0}Rez[ e^{iqr}q^{-1/2} \frac{q^2 I_2(lq)}{I_1(lq)}, q]
\ee
We get 
\bea
&&
\partial_y^2 \Phi
|_{y=0}\simeq -\frac{R_s l}{\sqrt{2\pi r}}\frac{2\pi \sqrt{l}}{l^2}
\sum_n j_{1,n}^{-1/2}\frac{j_{1,n}^2 J_2(j_{1,n})}{l a_{1,n}}e^{-\bar{q}_n r}
\nonumber\\&&
\simeq -\frac{R_s}{l^2} \sqrt{\frac{2\pi l}{r}}\tilde{C}_1 e^{-\bar{q}_1r}
= -\Phi |_{y=0}\bar{q}_1^2
\eea
(since $\tilde{C}_1=C_1(j_{1,1})^2$). 

We are now ready to apply the formalism for the trapped surface in curved 
background. In
\be
\Psi= f+a y +g\frac{y^2}{2}+...=C
\ee
we have
\be
a=\zeta_1, \;\;\; g= \partial_y^2\Phi|_{y=0}-\frac{4}{l} \zeta_1,\;\;\ 
f= \Phi |_{y=0}
\ee
We  first try $a=\zeta_1=0$ (so that $\Psi= \Phi$). Then we need to 
match 
\bea
&&C= f+g\frac{y^2}{2}+...\nonumber\\&&
4=f'^2 +y^2(f'g'+g^2)+...
\eea
where $g= -f \bar{q}_1^2$ and $f'=-f\bar{q}_1$. We obtain 
\bea
&& C^2=f^2(a-y^2 \bar{q}_1^2)+...\nonumber\\&&
4=f^2 \bar{q}_1^2(1+0)+...
\eea
which we see that don't match. So we need to put a nonzero $a=\zeta_1$.

Then we would need to match 
\bea
&& C=f+ay+...\nonumber\\&&
4=f'^2 +a^2 + y (2a'f'+2\frac{a^2}{l}+2ag)+...
\eea
where $f'=-\bar{q}_1f$ and $g= -\bar{q}_1^2 f -4 a/l$. 

We will try first $a=\alpha f/l$, which gives a term in the 
continuity equation comparable with the leading term. We get 
\bea
&&C^2=f^2 (1+2\frac{\alpha}{l}y)+...\nonumber\\
&&\frac{4}{\bar{q}_1^2+\alpha^2/l^2}= f^2[1+ \frac{y}{\bar{q}_1^2+\alpha^2/l^2}
(-\frac{6\alpha^2}{l^3})]+...
\eea
and by matching the two equations we get 
\be
\alpha = \frac{-3\pm \sqrt{9-4l^2\bar{q}_1^2}}{2}
\ee
however, since $j_{1,1}=\bar{q}_1l\simeq 3.83$, there is no real solution!

Next, we try an order of $l/r$ down from the previous try, namely 
$a=\beta f/r$. We get 
\bea
&&C^2 =f^2(1+2\frac{\beta}{r}y)+...\nonumber\\&&
4= f^2\bar{q}_1^2(1+o(\frac{yl}{r^2}))+...
\eea
so now even the r dependence doesn't match. We can see that by adding powers 
of $l^n/r^n$ we generate again a mismatch of r dependence. 

Conversely, we can try to have an $a=\zeta_1$ which is more important than the 
leading term, namely $a=\alpha fr/l^2$. Then we get 
\bea
&& C^2=f^2(1+2\alpha\frac{yr}{l^2})+...\nonumber\\&&
4=f^2\frac{\alpha^2r^2}{l^4}(1-\frac{6y}{l})+...
\eea
so again a mismatch of r dependence. 

For higher powers of $r^n/l^n$
\be
a=\alpha \frac{f\;r}{l^2}(\frac{r}{l})^n
\ee
we need to match 
\bea
&&C^2=f^2[1+2\alpha \frac{yl}{r^2}(\frac{r}{l})^n]
\nonumber\\
&&4=\frac{f^2 \alpha^2 r^2}{l^4}[1-\frac{6y}{l}(\frac{r}{l})^{2n}]
\eea
and we see that we get a solution for $n=1, \alpha =-3$, thus 
\be
a= -3\frac{f \; r^2}{l^3}
\ee

It would seem that we have exhausted all the possibilities, but this is 
actually not so. We can still try 
\be
a=\zeta_1= \alpha \frac{e^{\beta r}}{r}f
\ee
which gives
\bea
&&C^2=f^2 (1+2\alpha \frac{ye^{\beta r}}{r})+...\nonumber\\&&
4=f^2 \bar{q}_1^2[1-2\alpha \frac{ye^{\beta r}}{r\bar{q}_1^2}
(\beta \bar{q}_1+3\alpha \frac{e^{\beta r}}{lr})]+...
\eea
we can easily see that $\beta =-\bar{q}_1$ makes the two equations equal, so 
\be
a=\zeta_1= \alpha \frac{e^{-\bar{q}_1 r}}{r}f
\ee
and strangely $\alpha $ is arbitrary (but probably  it will be fixed 
in a higher order in y). 

In any case, we see that this solution for $a$ does not change the 
leading order equation at y=0, which is 
\be
\frac{|f|\bar{q}_1}{2}=1\Rightarrow \frac{R_s\bar{q}_1}{2}
\sqrt{\frac{2\pi l}{r}}C_1 e^{-\bar{q}_1r}=1 
\ee
which has a solution that is approximately 
\bea
&&r_H\simeq \frac{1}{2\bar{q}_1}ln (2\bar{q}_1 A)\nonumber\\
&&A= (\frac{R_s\bar{q}_1}{2})^2 2\pi l C_1^2 
\eea

The fact that there is a solution can be easily seen by considering the 
function
\be
\tilde{g}(r)= r-A e^{-2\bar{q}_1r}\Rightarrow \tilde{g}'(r)= 1+ 
2\bar{q}_1 A e^{-2\bar{q}_1r}>0
\ee
and the solution we are looking for is given by $\tilde{g}(r)=0$.  
Since $\tilde{g}(0)=-A$ and the function is monotonically increasing, it 
will have a solution. 

At nonzero impact parameter, we get the equation
\be
(\frac{f\bar{q}_1}{2})^2 (1-\frac{b^2}{2r^2})=1
\ee
Its  solution is thus the zero of the function
\be
g(r)= r- (1-\frac{b^2}{2r^2})Ae^{-2\bar{q}_1r}
\ee
but now the analysis of the solution is a bit more involved. Indeed, now
\be
g'(r)= 1+2\bar{q}_1 A(1-\frac{b^2}{2r^2})e^{-2\bar{q}_1r}-\frac{b^2}{r^3}Ae^{
-2\bar{q}_1r}
\label{derivative}
\ee
Luckily, we are only interested in the maximum value $b_{max}$ 
for which $g(r)=0$ has a solution. 

If $r^2<b^2/2$, then $g(r)>0$. In particular, $g(0)=+\infty$, $g(b/\sqrt{2})
= b/\sqrt{2}$. To have a solution of $g(r)=0$, 
we need a minimum, $g'(r_1)=0$, with 
$g(r_1)<0$, so necessarily $r_1^2>b^2/2$.

But if $1-b^2/(2r^2)\sim 1$ ($b^2/2$ significantly lower than $r^2$), 
the second term in (\ref{derivative}) is larger than the third, so 
$g'(r)>0$ (as $\bar{q}_1\sim 1/l\gg 1/r$). Thus we need instead
\be
\frac{b^2}{2r_1^2}=1-\epsilon\simeq 1
\ee
so if $g(r_1)<0$ we get 
\be
g(r_1)\simeq r_1-\frac{A}{\bar{q}_1r_1}e^{-2\bar{q}_1r_1}
\ee

At $b=b_{max}$,  $g(r_1)=0$, so 
\bea
&&r_1=\frac{A}{\bar{q}_1r_1}e^{-2\bar{q}_1r_1},\;\; b^2\simeq 2r_1^2
\nonumber\\&&
\Rightarrow
b_{max}(\sqrt{s})\simeq \sqrt{2} \frac{1}{2\bar{q}_1}ln (4 \bar{q}_1A)=
 \frac{\sqrt{2}}
{\bar{q}_1}ln [ R_s \bar{q}_1(\sqrt{j_{1,1}}C_1\sqrt{
\frac{\pi}{2}})\sqrt{2}]
\eea
where $R_s=\sqrt{s}G_4$, $G_4= 1/(l M_{P,5}^3)$.

However, the same physical argument we have used in the AdS case applies.
The two solutions we have obtained correspond to the trapped surface that 
would be there if we had a four dimensional scattering, and only the 
function $\Phi$ would be different (for $a=\alpha e^{-\bar{q}_1 r}f/r$), 
and the one due to the very large warping outside the 4d IR brane. 
As the large warping will increase the size of the black hole being 
formed, we have to take the larger trapped surface, described by 
\be
a=-3\frac{f \; r^2}{l^3}
\ee
which gives the continuity condition at y=0
\be
(\frac{|f|\bar{q}_1}{2})\frac{|\alpha |r}{\bar{q}_1l^2}=1
\Rightarrow \frac{R_s\bar{q}_1}{2}
\sqrt{\frac{2\pi l}{r}}\frac{|\alpha |C_1}{\bar{q}_1 l}\frac{r}{l}
 e^{-\bar{q}_1r}=1 
\ee
which has a solution that is approximately
\bea
&&r_H\simeq \frac{1}{2\bar{q}_1}ln (\frac{\bar{A}}{2\bar{q}_1})\nonumber\\
&& \bar{A}= (\frac{R_s \bar{q}_1}{2})^2 \frac{2\pi }{l} (\frac{3C_1}{
\bar{q}_1 l})^2 
\eea

We can again see that there is a solution by considering the function
\be
\tilde{g}(r)= r-\bar{A}r^2 e^{-2\bar{q}_1r}\Rightarrow 
\tilde{g}'(r)= 1+(\bar{q}_1-\frac{1}{r})2\bar{A}r^2 e^{-2\bar{q}_1r}
\ee
and the solution for $r_H$ is given by $\tilde{g}(r)=0$. We can see that 
$\tilde{g}'(r)>0$ if $r>1/\bar{q}_1$ and 
\be
\tilde{g}(\frac{1}{\bar{q}_1})= \frac{1}{\bar{q}_1}[1-\frac{R_s^2}{l^2}
\frac{18 \pi C_1^2}{4 e^2 j_{1,1}}]<0
\ee
for sufficiently large $R_s$, so there will be a solution.

\end{document}